\documentclass[%
 aip,
 amsmath,amssymb,
reprint, onecolumn 
]{revtex4-1}

\usepackage{graphicx}
\usepackage{dcolumn}
\usepackage{bm}

\usepackage[utf8]{inputenc}
\usepackage[T1]{fontenc}
\usepackage{mathptmx}
\usepackage{etoolbox}

\usepackage{amsfonts}
\usepackage{amssymb}
\usepackage{appendix}
\usepackage{amsmath}
\usepackage[margin=0.8in]{geometry}
\usepackage{ esint }
\usepackage{enumitem}
\usepackage{physics}
\usepackage{tikz}
\usepackage{pgfplots}
\usepackage{dsfont}
\usepackage{dutchcal}
\usepackage{babel}
\usepackage{marginnote}
\usepackage{relsize}

\newtheorem{thm}{Theorem}
\newtheorem{prop}[thm]{Proposition}
\newtheorem{lem}[thm]{Lemma}
\newtheorem{cor}[thm]{Corollary}
\newtheorem{dfn}[thm]{Definition}

\numberwithin{equation}{section}

\newtheorem{Rem}{Remark}

\makeatletter
\def\@email#1#2{%
 \endgroup
 \patchcmd{\titleblock@produce}
  {\frontmatter@RRAPformat}
  {\frontmatter@RRAPformat{\produce@RRAP{*#1\href{mailto:#2}{#2}}}\frontmatter@RRAPformat}
  {}{}
}%

\newcommand{\eff}{{\mathrm{eff}}}
\makeatother

\begin{document}

\title[On the localization regime of certain random operators within Hartree-Fock theory]{On the localization regime of certain random operators within Hartree-Fock theory}
\author{Rodrigo Matos\\Department of Mathematics, PUC-Rio, 22451-900, Rio de Janeiro, Brasil}

 \altaffiliation[rodrigo@mat.puc-rio.br]{}

\date{\today}

\begin{abstract}
Localization results for a class of random Schr\"odinger operators within the Hartree-Fock approximation are proved in two regimes: large disorder and weak disorder/extreme energies. A large disorder threshold $\lambda_{\mathrm{HF}}$ analogous to the threshold $\lambda_{\mathrm{And}}$  obtained in [J. Schenker, Lett. Math. Phys, Vol 105, 1 (2015)] is provided. We also show certain stability results for this large disorder threshold by giving examples of distributions for which $\lambda_{\mathrm{HF}}$ converges to $\lambda_{\mathrm{And}}$, or to a number arbitrarily close to it, as the interaction strength tends to zero.
\end{abstract}
\maketitle

\section {Introduction}
In recent decades there has been intense activity regarding mathematical aspects of disordered systems. In the Anderson model in dimension two or higher, there is an extensive literature regarding localization in the regimes of large disorder or at spectral edges. In this context, proofs either follow the strategy of the multiscale analysis, see \cite{Bourgain-Kenig,Carm-Klein-Mart,Ding-Smart,Fr-M-S-S85,Fr-Spenc,Germ-Hislop-Klein,GKBootstrap,Kir-Stol-Stolz,vonD-K} 
and also the surveys \cite{Kirschsurv,Klein-surv}, or the method of fractional moments, dating back to \cite{A-Molc,Aiz-weakdis} and further developed in both discrete and continuous settings \cite{A-S-F-H,Aiz-Elg-N-S-S} and also in the context of non-monotone potentials \cite{Elg-Taut-Ves,E-S-S}, see also the survey \cite{Stolz-surv} and the monograph \cite{A-W-B}. Localization in the context of weak disorder and existence of the so-called Lisfshitz tails were also extensively studied, see \cite{Aiz-weakdis,Elgart-LT3d, RojasM-Geb1,RojasM-Geb2,Klopp-Lif,Klopp-Nak,Wang-univ} and references therein. For results on complete localization in one dimension using large-deviation techniques we refer to \cite{Jit-Z,Bucaj1dloc}.\par
In the past years, there has been a number of developments in the context of many-body disordered systems such as systems with a finite number of particles \cite{A-W-Mult, Chul-Suh,Chu-Suh-ind}; the quantum $XY$ \cite{AR-N-S-S,H-S-Stolz,Klein-Perez}
and $XXZ$ \cite{E-K22,E-K-S-2,E-K-S-1} spin chains; systems of hardcore particles \cite{Beaud-W}; harmonic oscillators in the presence of disorder \cite{N-S-S-harmonic}; particle-oscillator interactions \cite{S-M-Holst}. Unlike in the single-particle Anderson-type models, where the notions of localization aimed at are usually spectral or dynamical localization, the challenges in the context of true many-body quantum systems start at defining the correct objects and notions of localization for each model.
\par One alternative to explore interactive quantum systems while remaining closer to the single-particle Schr\"odinger operator setting is to approximate the true many-body Hamiltonian by an effective one, as in the case of mean field theories and the Hartree/Hartree-Fock approximations which are widely studied beyond the setting of disordered systems \cite{B-Lieb-S,Bene-Porta-Schlein,Canc-Lab-Lew,CHENN2022109702,Hainzl-Lew-Spa,Lah-1,Lieb-Simon}.\par

 In the disordered setting, Anderson localization in the Hartree-Fock approximation was first studied in \cite{Duc}. There, through the multiscale analysis technique, spectral localization was obtained in the presence of a spectral gap at both large disorder and at spectral edges. Recently in \cite{M-S}
 localization properties of the disordered Hubbard model at positive temperature within the Hartree-Fock approximation have been  established via the Aizenman-Molchanov fractional moment technique. There, exponential dynamical localization ( in fact, decay of eigenfunction correlators) is shown to hold at large disorder in dimension $d\geq 2$ and at any disorder in dimension $d=1$ provided the interaction strength is sufficiently small. No assumption on the existence of a spectral gap is made but, in contrast, the interactions are modelled at positive temperature. 
 The present manuscript is devoted to localization properties of random operators in the form \begin{equation}\label{toymodel}
     H_{\omega}=-\Delta+\lambda V_{\omega}+gV_{\eff,\omega}
\end{equation}
where $\{V_{\omega}(n)\}_{n\in \mathbb{Z}^d}$ are independent, identically distributed random variables and $V_{\eff,\omega}$ is a multiplication operator implicitly defined by
\begin{equation}\label{effpot}
     V_{\eff,\omega}(n)=\sum_{m\in \mathbb{Z}^d}a(n,m)\langle \delta_m,F(H_{\omega})\delta_m\rangle\, \text{for all}\, n\in \mathbb{Z}^d.
\end{equation}
Here $|a(m,n)|\leq C_ae^{-\gamma d(m,n)}$ will be assumed to decay sufficiently fast ( see \ref{assump1}-\ref{assump7} for the precise assumptions) with respect to a metric $d:\mathbb{Z}^d\times \mathbb{Z}^d\rightarrow \mathbb{R}$, $C_a>0$, $\gamma>0$ and $F$ is an analytic function on a strip $\{|\mathrm{Im}z|<\eta\}$ which is bounded. It is worth noting that the above setting allows for the decay of $|a(m,n)|$ to be of polynomial type. The above model is somewhat analogous (in the Hartree-Fock setting) to models in the single-particle setting with fast decaying potentials which still exhibit monotonicity properties (for instance, the ones studied in \cite{Kir-Stol-Stolz}).
In the particular case where
$F(z)=\frac{1}{1+e^{\beta(z-\bar \mu)}}$ is the Fermi-Dirac function at temperature $\beta^{-1}>0$ and a chemical potential $\bar\mu\in \mathbb{R}$ and $a(m,n)=\delta_{mn}$ ( with $\delta_{mn}$ the Kronecker delta) \eqref{effpot} simplifies to operators already studied in \cite{M-S}. There  for a fixed $\beta>0$, dynamical localization is shown in any dimension provided $|g|<g_0$ and $\lambda>\lambda_0$ for certain constants $g_0$ and $\lambda_0$ which depend on $\beta$ and $d$ but a more concrete estimate for $\lambda_0$ was not pursued there. In this note, we generalized the large disorder result of \cite{M-S} to the operator \eqref{effpot}, obtain a novel result of localization at weak disorder/extreme energies and, moreover, study the question of stability of the large disorder threshold under `weak' interactions, inspired by the analysis of \cite{Schenkl}. In particular, it is proven here that there is a  large disorder threshold $\lambda_{\mathrm{HF}}$ such that the operators given by \eqref{effpot} exhibit dynamical localization provided  $\lambda>\lambda_{HF}$ as long as $|g|\norm{F}_{\infty}$ is sufficiently small. Moreover, we show that
$\lambda_{HF}\to \lambda_{\mathrm{And}}$ as $|g|\norm{F}_{\infty} \to 0$ where $\lambda_{\mathrm{And}}$ is the solution of the transcendental equation \begin{equation}\label{eq:transced}\lambda_{\mathrm{And}}=2\norm{\rho}_{\infty}\mu_d e\ln\left(\frac{\lambda_{\mathrm{And}}}{2\norm{\rho}_{\infty}}\right)
\end{equation}
with $\mu_d$ the connectivity constant of $\mathbb{Z}^d$. For the uniform distribution in $[-1,1]$, in which case $2\norm{\rho}_{\infty}=1$, this value of $\lambda_{\mathrm{And}}$ was obtained for the Anderson model in \cite{Schenkl} and coincides with Anderson's original prediction in \cite{Anderson}. To the best of our knowledge, in arbitrary dimension $\lambda_{\mathrm{And}}$ in \eqref{eq:transced} is the best rigorous large disorder threshold proved with current methods. It is worth noting that letting $|g|\norm{F}_{\infty} \to 0$ formally in \eqref{effpot} we obtain the Anderson model $H_{\mathrm{And}}=-\Delta+V_{\omega}$. \par
We now comment on other technical merits of the present work, for further technical aspects we refer to section \ref{sec:strategy} below. Our first observation is that even for the non-interacting Anderson model $H_{\mathrm{And}}$, the fractional moment method requires the random variables $V_\omega$ to have a density $\rho$ which is ``sufficiently regular". Thus, it is to be expected that a direct application of this technique to interacting models, which is the approach adopted here and also in \cite{M-S}, will require further regularity of $\rho$.
The previous paper \cite{M-S} covers a large class of probability distributions with $\mathrm{supp}\rho=\mathbb{R}$ by making use of the symmetry $F(z)=1-F(-z)$ and decay properties of the Fermi-Dirac function when $\mathrm{Re} z \to \infty$ in order to obtain certain improved Combes-Thomas bounds. Such bounds reflect decay of the effective potential at a given site $V_{\eff,\omega}(n)$ when the local potential $\omega(m)$ is changed at a site $m\neq n$. However, such bounds appear not to be available in the generality studied here. In fact, they seem not to be available even when one restricts \eqref{effpot} to the case of nearest neighbor lattice fermions, i.e., when
$a(m,n)=1$ if and only if $|m-n|=1$ where $|m|=|m_1|+\cdots+|m_d|$ and thus
$V_{\eff,\omega}(n)=\sum_{n'\sim n}\langle \delta_{n'},F(H_{\omega})\delta_{n'}\rangle$ with $n'\sim n$ indicating that $n'$ and $n$ are nearest neighbors on $\mathbb{Z}^d$.
The key observation surrounding the present paper is that there is a trade off between the regularity/decay properties of $F$ on the real line, the decay properties of the interaction kernel $a(m,n)$ and the density $\rho$. Namely, by reducing the class of probability distributions covered by our main result, we are able to include interactions of a much longer range, including the case where $a(m,n)$ only decays in an algebraic fashion and where $F(z)$ is bounded of a strip but does not necessarily decay as $\mathrm{Re}z\to \infty$. In conclusion, even though the methods employed here to obtain the \emph{a-priori} bound on fractional moments of the Green's function follow the general scheme of \cite{M-S}, in order to prove our stability result, we need to keep an explicit dependence on all parameters involved $\eta,\lambda,g,\norm{F}_{\infty}$ and now have the inclusion the decay rate $\gamma$ of $a(m,n)$ as well. Once an \emph{a-priori} bound on fractional moments of the Green's function is obtained, we follow the approach of Schenker in \cite{Schenkl} in order to get the best large disorder threshold which seem to be available with current methods which turns out to converge to $\lambda_{\mathrm{And}}$ when $|g|\norm{F}_{\infty} \to 0$.\par
This paper is dedicated to Abel Klein in occasion of his 78th birthday. Klein's contributions to the field go well beyond the aforementioned works and can hardly be overstated. Certain aspects of the present work were also inspired by Klein's efforts. 
For instance, the idea of studying distributions near a suitable chosen density (for which explicit calculations are available) used below in assumption \ref{assump6} is analogous to \cite{Acosta-Klein}, where analyticity of the density of states on a strip is shown for distributions sufficiently close to the Cauchy distribution. Moreover, throughout the note, Combes-Thomas type bounds for kernels of analytic functions of $H_{\omega}$ are used extensively. 
In \cite{GK-CT} such bounds are obtained in great generality which provides hope for future extensions of the results below.

\section{Model, Statement of the main results and proof strategies}
This note concerns random operators of the form \begin{equation}\label{toymodel2}
     H_{\omega}=A+\lambda V_{\omega}+gV_{\eff,\omega}
\end{equation}
acting on $\ell^2\left(\mathbb{Z}^d\right)$ as follows: 
\begin{enumerate}
\item \label{assump1} $(A\psi)(n)=\sum_{|n'-n|=1}\psi(n')$ for each $\psi \in \ell^2\left(\mathbb{Z}^d\right)$, i.e. $A:\ell^2\left(\mathbb{Z}^d\right)\rightarrow \ell^2\left(\mathbb{Z}^d\right)$ is the adjacency operator of $\mathbb{Z}^d$.
\item \label{assump2} $(V_{\omega}\psi)(n)=\omega(n)\psi(n)$ for each $\psi \in \ell^2\left(\mathbb{Z}^d\right)$ where $\{\omega(n)\}_{n\in \mathbb{Z}^d}$ are independent, identically distributed random variables with a bounded density $\rho$.
\item \label{assump3}
 The effective potential $V_{\eff,\omega}:\ell^2\left(\mathbb{Z}^d\right)\rightarrow \ell^2\left(\mathbb{Z}^d\right)$ is a multiplication operator implicitly defined by
\begin{equation}\label{effpot2}
     V_{\eff,\omega}(n)=\sum_{m\in \mathbb{Z}^d}a(n,m)\langle \delta_m,F(H_{\omega})\delta_m\rangle\,\,\, \text{for all}\,\,\, n\in \mathbb{Z}^d.
\end{equation}
We impose the following conditions on $a(n,m)$ and $F$.

\item \label{assump4} There exists $\eta>\eta_0>0$ such that $F$ is an analytic function on the strip $$\mathcal{S}_{\eta}=\{|\mathrm{Im}z|<\eta\}$$ and bounded on its closure $\overline{\mathcal{S}_{\eta}}$. Moreover, we assume that $F(\mathbb{R})\subset \mathbb{R}$.
\item \label{assump5} The values $a(m,n)$ are real numbers for all $m,n\in \mathbb{Z}^d$ and 
\begin{equation}
 |a(m,n)|\leq C_ae^{-\gamma_ad(m,n)}   
\end{equation} for constants $C_a>0$ and $\gamma_a>0$ and some metric $d:\mathbb{Z}^d\times \mathbb{Z}^d \rightarrow \mathbb{R}$ for which there exists $\delta \in (0,\gamma_a/2)$ such that
\begin{equation}
\norm{S_{\delta-\gamma_a}}_{\infty,\infty}:=\sup_{n\in \mathbb{Z}^d}\sum_{m\in \mathbb{Z}^d}e^{(\delta-\gamma_a)d(m,n)}<\infty.
\end{equation}

    \item \label{assump6} We also assume that $\mathrm{supp}\rho =\mathbb{R}$ and that for some $c_1>0$ and $\varepsilon_1>0$  \begin{equation}\label{eq: fluctuation1ass}
    \frac{\rho(v_1)}{\rho(v_2)}\geq e^{-c_1|v_1-v_2|},\,\,\,\text{for all}\,\,v_1,v_2\in \mathbb{R}
    \end{equation}
    and \begin{equation}\label{eq:fluctassumption}
        \sup_{v\in \mathbb{R}}\frac{\rho(v)}{\int^{\infty}_{-\infty}\rho(\alpha)e^{-\varepsilon_1\abs{v-\alpha}}\,d\alpha}<\infty
    \end{equation}
    \end{enumerate}
    \begin{Rem}
    Assumptions \ref{assump1}-\ref{assump6} suffice for the first result of this note, namely, localization at large disorder given by Theorem \ref{thm:largedisorder} below, and also for the stability bounds on the large disorder threshold of Corolary \ref{cor:stability}. It is worth observing that assumption \ref{assump6} holds, for instance, for the Cauchy distribution and also for the (negative) exponential distribution.
    \end{Rem}
    For the results of localization at weak disorder/extreme energies we will make the following additional requirement.
    
\begin{enumerate}[resume,]
    \item \label{assump7} We further assume that $\rho(v)=h(v)e^{-c_{\rho}|v|}$ for some $c_{\rho}>0$ where
    \begin{equation}\label{eq: fluctuation1ass2}
    \frac{h(v_1)}{h(v_2)}\geq e^{-\varepsilon_2|v_1-v_2|} \,\,\,\text{for all}\,\,v_1,v_2\in \mathbb{R}
    \end{equation}
    for some $\varepsilon_2 \in (0,\frac{1}{2}c_{\rho})$. 
\end{enumerate}
\begin{Rem}
Intuitively speaking, assumption \ref{assump7} means that $\rho(v)$ is near the (negative and two sided) exponential distribution with density $\rho_1(v)=\frac{c_{\rho}}{2}e^{-c_{\rho}|v|}$ and $\mathrm{supp}\rho_1=\mathbb{R}$. With the help of the mean value theorem and Young's inequality, one may check that this assumption applies to densities of the form $\rho(v)=h(v)e^{-c|v|}$ with $h(v)=C_{k,\varepsilon}(1+\varepsilon|v|^k)e^{-\alpha|v|}$, $\alpha>0$, $k>1$ and $\varepsilon$ sufficiently small depending on $\alpha$ and $k$.
\end{Rem}
 Working with finite volume restrictions of both $H_{\omega}$ and also $V_{\eff,\omega}$ will turn out convenient thus we let $\Lambda_L=[-L,L]^d\cap \mathbb{Z}^d$ and
\begin{equation}\label{finitevoleffpot}
     V_{\eff,\omega,L}(n)=\sum_{m\in \Lambda_L}a(n,m)\langle \delta_m,F(H_{\omega,L})\delta_m\rangle\,\,\, \text{for all}\, n\in \Lambda_L.
\end{equation}
where $H_{\omega,L}=\mathds{1}_L (A+V_{\omega}+ V_{\eff,\omega,L})\mathds{1}_L$ and 
$\mathds{1}_L:\ell^2\left(\mathbb{Z}^d\right)\rightarrow \ell^2\left(\Lambda_L\right) $ is the projection onto $\mathrm{span}\{\delta_l:\,\,\,l\in \Lambda_L\}$.
 We will often write
\begin{equation}
U(n)=\omega(n)+\frac{g}{\lambda}V_{\eff,\omega,L},\,\,\,n\in \Lambda'\subset \Lambda_L
\end{equation}
 to denote the ``full potential" at site $n$. It will be shown below in Lemma \ref{lem:linfinitycontrol} that under assumptions \ref{assump1}-\ref{assump6} the conditional distribution of $U(n_0)=v$ at specified values of $\{U(n)\}_{n\in \Lambda'\setminus\{ n_0\}}$
has a density which is bounded with an upper bound independent of the parameters $\omega,\Lambda'$ and $L$. This upper bound is denoted herein by $M_{\infty}$ and the conditional density by $\rho^{\eff}_{n_0}=\rho^{\eff,\Lambda'}_{n_0,L}$.
We also recall the definition of the eigenfunction correlators for an operator $H$:
\begin{equation} Q_{I}(m,n):=\sup_{|\varphi|\leq 1}\abs{\langle \delta_m,\varphi(H)\delta_n\rangle}
\end{equation}
where the supremum is taken over Borel measurable functions $\varphi$ bounded by one and supported on the interval $I$. In case $I=\mathbb{R}$ we simply write $Q(m,n)$.
In what follows we denote by $Q^{\Lambda'}_{I,L}(m,n)$ the eigenfunction correlators of $H^{\Lambda'}_{\omega,L}={\mathds{1}}_{\Lambda'}H_{\omega,L}{\mathds{1}}_{\Lambda'}$ for $\Lambda'\subset \Lambda_L$ and by $\mathbb{E}(f)$ the expected value of $f$ with respect to the probability space in question. 

Our first result is the following.
\begin{thm}\label{thm:largedisorder}
 Under assumptions \ref{assump1}-\ref{assump6}
there exist $\lambda_{\mathrm{HF}}=\lambda_{\mathrm{HF}}(g,\eta_0,\norm{F}_{\infty},d,\rho,\gamma_a)$ and $g_0=g_0(C_a,d,\rho,\lambda,\gamma_a,\eta_0)$   such that for all $\lambda>\lambda_{\mathrm{HF}}$ and $\abs{g}\norm{F}_{\infty}<g_0$ we have that
\begin{equation}\label{eq:dynloc}
    \mathbb{E}\left(Q^{\Lambda'}_{L}(m,n)\right)\leq Ce^{-\nu'|m-n|}
\end{equation}
for some $\nu'>0$ and $C>0$ independent of $L$ and $\Lambda'$.
Moreover, $\lambda_{\mathrm{HF}}$ satisfies

\begin{equation}\label{eq:transc equation}
\lambda_{\mathrm{HF}}=2M_{\infty}\mu_d e\ln\left(\frac{\lambda_{\mathrm{HF}}}{2M_{\infty}}\right)
\end{equation}
where $\mu_d$ is the conective constant of $\mathbb{Z}^d$ and
$M_{\infty}=M_{\infty}(\eta_0,d)$ is given by

\begin{equation}
M_{\infty}=\sup_{\omega\in \Omega}\sup_{L\in \mathbb{N}}\sup_{\Lambda' \subset \Lambda_L}\sup_{n_0\in \Lambda}\sup_{v\in \mathbb{R}}\rho^{\eff,\Lambda'}_{n_0,L}(v).
\end{equation}
\end{thm}
\begin{Rem}
It readily follows that the analogue of \eqref{eq:dynloc} also holds in the infinite volume, see Lemma \ref{lem:normresconv} below and \cite[Proposition 7.6]{A-W-B}.
    
\end{Rem}

Theorem \ref{thm:largedisorder} above extends to the present context a result of Schenker  \cite{Schenkl}, who obtained the large disorder threshold $\lambda_{\mathrm{And}}$ which solves \begin{equation}\label{eq:transced2}\lambda_{\mathrm{And}}=2\norm{\rho}_{\infty}\mu_d e\ln\left(\frac{\lambda_{\mathrm{And}}}{2\norm{\rho}_{\infty}}\right).
\end{equation}
 for the Anderson model with a uniformly distributed potential on $[-1,1]$.\par
 We also show that
$\lambda_{\mathrm{HF}}$ is close to $\lambda_{\mathrm{And}}$ in a quantified fashion.

\begin{cor}\label{cor:stability}
Let $\lambda_{\mathrm{HF}}$ be as in \eqref{eq:transc equation} and $\lambda_{\mathrm{And}}$ be given by \eqref{eq:transced2}.
     Under assumptions \ref{assump1}-\ref{assump6} we have that
    $\abs{\lambda_{\mathrm{HF}}- \lambda_{\mathrm{And}}}\to 0$ as $|g|\norm{F}_{\infty} \to 0$.
\end{cor}

Before stating our second theorem we let, for each $n_0\in \Lambda'$
\begin{equation} \psi^{n_0}_s(z)=\int^{\infty}_{-\infty}\frac{|v|^s \rho^{\eff,\Lambda'}_{n_0,L}(v)}{|v-z|^s}\,dv,
\end{equation} 
\begin{equation}
\phi^{n_0}_s(z)=\int^{\infty}_{-\infty}\frac{\rho^{\mathrm{eff},\Lambda'}_{n_0,L}(v)}{\abs{v-z}^s}\,dv
\end{equation}
and 
\begin{equation}\label{eq:decoupconst}
    D_{s,1}=\sup_{L\in \mathbb{N}}\sup_{\Lambda'\subset \Lambda_L}\sup_{z\in \mathbb{C},n_0\in \Lambda'}\frac{\psi^{n_0}_s(z)}{\phi^{n_0}_s(z)}.
\end{equation}
As we shall see below, under assumptions \ref{assump1}-\ref{assump7} the measure $\rho^{\mathrm{eff},\Lambda'}_{n_0,L}(v)\,dv$ is $1$-moment regular in the sense of \cite[Definition 8.5]{A-W-B} meaning that $D_{s,1}<\infty$ for all $s\in (0,1)$.
We also define the Green's function of $H_{\omega}$ at $z\in \mathbb{C}\setminus \sigma(H_{\omega})$ by
\begin{equation}\label{eq:Green'sfunc}
G(m,n;z)=\langle \delta_m,(H_{\omega}-z)^{-1}\delta_n \rangle
\end{equation}
and let, for $\Lambda'\subset \Lambda_L$, $G_{L}(m,n;z)$ and $G^{\Lambda'}_{L}(m,n;z)$ be the Green's function of $H_{\omega,L}$ and $H^{\Lambda'}_{\omega,L}={\mathds{1}}_{\Lambda'}H_{\omega,L}{\mathds{1}}_{\Lambda'}$, respectively:
\begin{equation}
 G_L(m,n;z)=\langle \delta_m,(H_{\omega,L}-z)^{-1}\delta_n \rangle\,\,\text{and}\,\, G^{\Lambda'}_L(m,n;z)=\langle \delta_m,(H^{\Lambda'}_{\omega,L}-z)^{-1}\delta_n \rangle.
\end{equation}
We emphasize that the effective potential is $V_{\eff,\omega,L}$ for both of the above operators.
Finally, we denote by $G_0(m,n;z)$ the Green's function of the ``free'' operator $A$, namely
\begin{equation}\label{eq:freeGreen'sfunc}
G_0(m,n;z)=\langle \delta_m,(A-z)^{-1}\delta_n \rangle.
\end{equation}
We are now ready to state our second Theorem which yields localization at weak disorder/extreme energies provided the interaction strength is not too large relative to the remaining parameters.

\begin{thm} \label{thm:weakdisorder}
Given $I\subset \mathbb{R}$ there exist
$\lambda_0=\lambda_0(I)$ and $g_1=g_1(C_a,d,\rho,\lambda,\gamma_a,\eta_0)$   such that whenever $\abs{g}\norm{F}_{\infty}<g_1$ and $|\lambda|<\lambda_0$ we have that

\begin{equation}
    \mathbb{E}\left(Q^{\Lambda'}_{I,L}(m,n)\right)\leq Ce^{-\nu'|m-n|}
\end{equation}
for some $\nu'>0$ and $C>0$ independent of $\Lambda'\subset{\Lambda_L}$ and $L$.
Moreover,
we have that \begin{equation}
    \lambda_0=\sup_{s\in (0,1)}\sup_{\mu>0}\inf_{E\in I}\widehat{\lambda}_{s,\mu}(E)
\end{equation}
where
\begin{equation}
 \widehat{\lambda}_{s,\mu}(E)=\left( D_{s,1}\sup_{\delta\neq 0}\sup_{u\in \mathbb{Z}^d}\sum_{v\in \mathbb{Z}^d}|G_0(u,v;E+i\delta)|^se^{\mu|u-v|}\right)^\frac{-1}{s}.
\end{equation}
\end{thm}

\begin{Rem} By the Combes-Thomas bound \cite[Theorem 10.5]{A-W-B}, Theorem \ref{thm:weakdisorder} is applicable when $I\cap \sigma(A)=\emptyset$. In particular, since it was assumed that $\mathrm{supp}\rho=\mathbb{R}$, this yields a non-trivial result for all $\lambda \neq 0$. We choose the above formulation for general $I\subset \mathbb{R}$ for future reference, as in more general settings localization at weak disorder may be established away from the $\ell^1$ spectrum of the deterministic part of $H_{\omega}$, see  \cite[Theorem 10.4]{A-W-B} and comments therein.
\end{Rem}

\subsection{Proof strategy: discrete subharmonicty bounds}\label{sec:strategy}
The proofs of Theorems \ref{thm:largedisorder} and \ref{thm:weakdisorder} follow the general scheme of the Aizenman-Molchanov fractional moment method \cite{A-Molc,Aiz-weakdis} and further refinements of their technique, in particular the one in \cite{Schenkl}, combined with tools from \cite{M-S} (and a few technical improvements on it).
Their approach requires the random potential to be sufficiently regular (even though it allows for certain singularities) which is the case in the Anderson model $H_{\mathrm{And},\omega}=A+\lambda V_{\omega}$ given by assumptions \ref{assump1} and \ref{assump2}. The first difficulty in the present work is that the full random potential is of the form $U_{\omega}(n)=\omega(n)+\frac{g}{\lambda}V_{\eff,\omega}(n)$ thus
$U(n)$ and $U(m)$ are correlated for all values of $m$ and $n$ and, a-priori, their regularity is unknown. While correlations are not necessarily a problem for the fractional moment technique, as it is well-known and already stated in the Aizenman-Molchanov original work \cite{A-Molc}, in order to prove localization one needs at least some regularity on the conditional distributions of $U(n_0)$, for each $n_0$, when the remaining variables $\{U(n)\}_{n\neq n_0}$ are specified. Moreover, the involved bounds should be uniform in $n_0$. At an intuitive level, such requirement on the conditional distributions amounts to the variables $U(m)$ and $U(n)$ being less and less correlated as $|m-n|\to +\infty$ so that some of the regularity of $\omega(n_0)$ is persists in the conditional distribution of $U(n_0)$. The technical implementation of the above reasoning essentially consists of two main parts, each of them having of a few steps. The first part is completely deterministic and aims at showing that, in terms of the metric $d$ in which $\abs{a(m,n)}\leq C_a e^{-\gamma_a d(m,n)}$, the effective potential $V_{\eff,\omega}$ is a quasilocal function of the random variables $\{\omega(n)\}_{n\in \mathbb{Z}^d}$. The second part involves applying the fractional moment method in the spirit of \cite{Schenkl} once the regularity of the $\{U(n)\}_{n\in \Lambda'}$ is determined.\par
Before stating the main steps of the proof, let us remark that for simplicity we do not always mention finite-volume restrictions in this sketch. Nonetheless, their introduction is technically important for the arguments, as it will be clear later in the note. Moreover, each point of the outline below is carried out in the appropriate smallness regime (cf. Theorems \ref{thm:largedisorder} and \ref{thm:weakdisorder}).
\begin{enumerate}[label=(\roman*)]
    \item \label{step1} Step 1: Show that
    \begin{equation*}
\abs{\frac{\partial V_{\eff,\omega}(n)}{\partial \omega(l)}}\leq C_1 e^{-\delta d(n,l)}
\end{equation*}
holds for \emph{every} $\omega\in \Omega$, $n,l\in \Lambda'$ with $C_1$, $\delta>0$ independent of $\omega$ and $\Lambda'$. This will allow us to make the change of variables
$\omega(n)\mapsto U(n):=\omega(n)+\frac{g}{\lambda}V_{\eff,\omega}(n)$ and guarantee that the map $\omega \mapsto U$ is a diffeomorphism in $\mathbb{R}^{\abs{\Lambda'}}$ for each finite set $\Lambda'\subset \mathbb{Z}^d$.

\item \label{step2} Step 2: Fix $n_0\in \Lambda'$ and $\alpha\in \mathbb{R}$. Let $U_{\alpha}(n)=U(n)+(\alpha-U(n_0))\delta_{n_0}$ be a rank-one perturbation of $U$ at $n_0$
and define $\{\omega_{\alpha}\}_{n\in \Lambda'}$ to be such that 
$U_{\alpha}(n)=\omega_{\alpha}(n)+\frac{g}{\lambda} V_{\eff,\omega_{\alpha}}(n)$ for all $n\in \Lambda'$. Then for some $\delta>0$ and $C>0$ we have that
\begin{equation}
\abs{\omega(n)-\omega_{\alpha}(n)}\leq C_1|\alpha-U(n_0)|e^{-\delta d(n,n_0)}.
\end{equation}
This step, along with assumption \ref{assump6} will allow us to control fluctuations of the density $\rho$ which naturally appear when computing the conditional density $\rho^{\eff}_{n_0}.$

\item \label{step3} Step 3: Prove that
\begin{equation}
\abs{\frac{\partial V_{\eff,\omega}(n)}{\partial \omega(l)}-\frac{\partial V_{\eff,\omega_{\alpha}}(n)}{\partial \omega(l)}}\lesssim e^{-\delta\left(d(n,l)+d(n,n_0)\right)}.
\end{equation}
This step will help us control fluctuations in the Jacobian of the above change of variables which also appear in the expression for $\rho^{\eff}_{n_0}$.
\end{enumerate}
Once the above steps are completed, the second part of the proof makes use of probabilistic techniques.
\begin{enumerate}[resume,label=(\roman*)]
    
\item \label{step4} Step 4: Use the bounds from Steps 1-3 to conclude that under assumptions \ref{assump1}-\ref{assump6} the conditional density $\rho^{\eff}_{n_0}$ exists and is (uniformly) bounded. Moreover, under assumptions \ref{assump1}-\ref{assump7} conclude that  $\rho^{\eff}_{n_0}$ exhibits some additional regularity.

\item \label{step5} Step 5: Complete the proof using the fractional moment technique.
\end{enumerate}

 While the overall strategy outlined above is similar to the one in \cite{M-S} there are some key technical differences. Firstly, by obtaining the cancellation directly on \ref{step3} we are able to avoid having to bound the second
derivatives $\frac{\partial^2 V_{\eff,\omega}(n)}{\partial \omega(m)\partial \omega(l)}$ which shortens the proof quite a bit, especially for the model studied here where $a:\mathbb{Z}^d\times \mathbb{Z}^d \rightarrow \mathbb{R}$ may be non-local. Secondly, in Step 4 the observation that further regularity of $\rho^{\eff}_{n_0}$ can be obtained under assumption \ref{assump7}, which ultimately yields the localization at weak disorder/extreme energies result, is also new. A third difference is present in Step 5. Namely, while localization at large disorder was obtained in \cite[Theorem 2]{M-S}, in the case where $a(m,n)=\delta_{mn}$, the explicit dependence of the large disorder threshold on the remaining parameters is not given (although it can certainly be inferred from the proof). Here we provide a self-consistent equation for the large disorder threshold in \eqref{eq:transc equation}. Moreover we show that under assumption \ref{assump7} this threshold is somewhat sharp from the point of view of what is currently known for the Anderson model from \cite{Schenkl}. Indeed, within the class of exponential distributions, we show that the difference between the large disorder threshold $\lambda_{\mathrm{And}}$ of the non-interactive setting ( cf. \eqref{eq:transced2}) and $\lambda_{\mathrm{HF}}$ given by \eqref{eq:transc equation} can be made arbitrarily small when the interaction strength tends to zero.\par
Turning to the question of how to show the quasilocality bounds in Steps \ref{step1}-\ref{step3}, the following Lemma will be useful since $V_{\eff,\omega}$ and, by extension, its partial derivatives $\{\frac{\partial V_{\eff,\omega}(n)}{\partial \omega(l)}\}_{n,l\in \Lambda'}$ are only implicitly defined and hence the desired control of them can only be achieved via inequalities of self-consistent nature.

\begin{lem}\label{Lem:subhar}\cite[Theorem 9.2]{A-W-B}
Let $\mathbb{G}$ be a countable set and $K:\ell^{\infty}\left(\mathbb{G}\right)\rightarrow \ell^{\infty}\left(\mathbb{G}\right)$ be given by $(K\varphi)(n)=\sum_{u\in \mathbb{G}}K(n,u)\varphi(u)$ with $K(n,u)\geq 0$ and
\begin{equation}
\|K\|_{\infty,\infty}:=\sup_{n\in \mathbb{G}}\sum_{u\in \mathbb{G}}K(n,u)<1.
\end{equation} Let $W:\ell^{\infty}\left(\mathbb{G}\right)\rightarrow (0,\infty)$ and $\psi\in \ell^{\infty}\left(\mathbb{G}\right)$  be positive functions such that
\begin{equation}\label{subharassump}
 b_1:=\sum_{u\in \mathbb{G}}W(u)\psi(u)<\infty \,\,  \mathrm{and}\,\,
b_2:=\sup_{m\in \mathbb{G}}\sum_{u\in \mathbb{G}}\frac{W(u)}{W(m)}K(u,m)<1.
\end{equation}
Then, any $\varphi \in \ell^{\infty}\left(\mathbb{G}\right)$ which satisfies
\begin{equation*}
0\leq \varphi(n)\leq \psi(n)+(K\varphi)(n) \,\,\,\,\mathrm{for}\,\mathrm{all}\,\, n\in \mathbb{G}
\end{equation*}
also obeys the bound
\begin{equation}
    \sum_{n\in\mathbb{Z}^d}W(n)\varphi(n)\leq \frac{b_1}{1-b_2} \,\,\,\,\mathrm{for}\,\mathrm{all}\,\, n\in \mathbb{G}.
\end{equation}
\end{lem}
The first instance where Lemma \ref{Lem:subhar} is applied is in Step 1 with the choice
\begin{equation}
 \varphi_1(n)=\abs{\frac{\partial V_{\eff}(n)}{\partial \omega(l)}},\,\, W(n)=e^{\delta|n-l|},\,\, \delta=\min\{\nu,\gamma_a/2\},
\end{equation}
where $\nu$ is given below in \eqref{eq:CTbound} and $\gamma_a$ is as in \ref{assump5}.
To accomplish Step 2, Lemma \ref{Lem:subhar} is applied to
\begin{equation}
 \varphi_2(n)=\abs{\omega(n)-\omega_{\alpha}(n)}\delta_{n\neq n_0},\,\,\, W(n)=e^{\delta|n-l|}.  
\end{equation}
with $\delta$ as above.
In Step 3, Lemma \ref{Lem:subhar} is applied to 
\begin{equation}
 \varphi_3(n)=\abs{\frac{\partial V_{\eff}(n)}{\partial \omega(l)}-\frac{\partial V_{\eff,\omega_{\alpha}}(n)}{\partial \omega(l)}},\,\,\, W(n)=e^{\delta|n-l|}. 
\end{equation}

Finally, in Step 5 Lemma 4 is applied to different functions depending on whether we wish to
show decay of the Green's function in the large disorder or in the weak disorder/extreme energies regime. In the large disorder regime of Theorem \ref{thm:largedisorder}, thanks to an \emph{a-priori} bound which follows from Lemma \ref{lem:linfinitycontrol} below, Lemma \ref{Lem:subhar} is applied to a fixed $n\in \mathbb{Z}^d$ letting
\begin{equation}
 \varphi(m)=\sup_{\Lambda\subset \mathbb{Z}^d}\mathbb{E}\left(\abs{G^{\Lambda}(m,n;z)}^s\right),\,\,\, W(m)=e^{\nu'|m-n|},   
\end{equation}
for a suitable $\nu'>0$
and choosing
\begin{equation}
K(m,u)=\frac{2^{s}M^s_{\infty}}{\lambda^s}\delta_{|m-u|=1},\,\,\, \psi(m)=\frac{2^{s}M^s_{\infty}}{\lambda^s}\delta_{m,n}.
\end{equation}
In the regime of weak disorder/extreme energies of Theorem \ref{thm:Greendecayweakdis}, Lemma \ref{Lem:subhar} can be applied to
  
 \begin{equation}
K(m,u)=D_{s,1}\abs{\lambda}^s\abs{G_0(m,u;z)}^s,\,\,\,\psi(m)=\abs{G_0(m,n;z)}^s
 \end{equation}
 thanks to Lemma \ref{lem:expdecaycontrol} below which implies a decoupling estimate for the Green's function fractional moments cf. \cite[Theorems 8.7 and 10.4]{A-W-B}\par
 The remainder of this note is organized as follows: in Section \ref{Sec:decaybounds} we show the quasilocality bounds of Steps 1 and 2 above, in Section \ref{Sec:canceldecay} we show the cancellation bound of Step 3, in Section \ref{Sec:Lemmas} we state and prove the technical Lemmas on the conditional densities $\rho^{\eff}_{n_0}$.The proofs of Theorems   \ref{thm:largedisorder} and \ref{thm:weakdisorder} as well as Corollary \ref{cor:stability} are given in Sections \ref{sec:largedisproofs} and \ref{Sec:weakdisproof}. In the Appendix we provide some basic facts about existence of the effective potential and norm resolvent convergence of finite volume restrictions to the infinite volume operator.
\section {First order decay bounds on the effective potential}\label{Sec:decaybounds}
Let us collect some basic facts which will be repeatedly used in this note. Firstly, if $H_{\omega,L}$ is as above
we can write $F(H_{\omega,L})=\frac{1}{2\pi i}\int^{\infty}_{-\infty}\left(\frac{1}{H_{\omega,L}+t-i\eta}-\frac{1}{H_{\omega,L}+t+i\eta}\right) f(t)\,dt$ for  $ f=F_{+} + F_{-} + D\ast F $, where $F_{\pm}(u)=F(u\pm i\eta\mp i0)$ and  $D(u)=\frac{\eta}{\pi\left( \eta^2+u^2\right)}$ the Poisson kernel, see \cite[Appendix D]{A-G}. In particular, the inequality $\|f\|_{\infty}\leq 3\|F\|_{\infty}$ holds. The formula
\begin{equation}\label{eq:inteff}
V_{\eff,\omega,L}(n)=\frac{1}{2\pi i}\int^{\infty}_{-\infty} K_L(n,\omega;t)f(t)\,dt
\end{equation}
with \begin{equation}\label{eq:kernel}
    K_L(n,\omega;t)=\sum_{m\in \mathbb{Z}^d}a(n,m)\left(G_L(m,m;t-i\eta)-G_L(m,m;t+i\eta)\right)
\end{equation} readily follows and is a useful representation for the effective potential. It is shown below that it yields, for each $n,l\in \Lambda_L$, self-consistent equations for the derivatives $\frac{\partial V_{\eff,\omega,L}(n)}{\partial\omega(l)}$ which in turn imply the desired exponential decay in Step 1 of the proof strategies given earlier. We introduce
$\nu>0$ such that 
\begin{equation}\label{eq:definitionnu}
\sup_{n\in \mathbb{Z}^d}\sum_{|n'-n|=1}e^{\nu d(n,n')}<\eta/2.
\end{equation}
The decay rate in the Lemma below will be dictated by $\nu$ and $\gamma_a$.
  \begin{lem}\label{lem:derivativedecay}
  Let $\nu>0$ be as in \eqref{eq:definitionnu} and $\gamma_a$ as in Assumption \ref{assump5}.
  For each $L\in \mathbb{N}$, $l\in  \Lambda_L=[-L,L]^d\cap \mathbb{Z}^d$ and any $\delta<\min\{\gamma_a,2\nu\}$ the inequality
  \begin{equation}\label{eq:Derivativebound}
  \sum_{n\in \Lambda}e^{\delta d(n,l)}\abs{\frac{\partial V_{\eff,\omega,L}(n)}{\partial\omega(l)}}\leq C_1
  \end{equation}
  holds whenever
  $\frac{C_a72\sqrt{2}\norm{F}_{\infty}}{\eta}|g| S_{\delta-\gamma}S_{\delta-2\nu}<\frac{1}{2}$, with $d:\mathbb{Z}^d\times \mathbb{Z}^d\rightarrow \mathbb{R}$ as in assumption \ref{assump5}, \begin{equation}\label{eq:C1}
  C_{1}=\lambda\frac{C_a144\sqrt{2}\norm{F}_{\infty}}{\eta} S_{\delta-\gamma}S_{\delta-2\nu},
  \end{equation}
  and \begin{equation}\label{eq:defexpsums}
    S_{\beta}:=\sup_{u\in \mathbb{Z}^d}\sum_{v\in \mathbb{Z}^d}e^{\beta d(u,v)}.  
  \end{equation}
  
  \end{lem}
 
  \textbf{Proof.}

Denote by $P_l:\ell^2\left(\mathbb{Z}^d\right)\rightarrow \mathrm{Span}\{\delta_l\}$ the projection onto $\mathrm{Span}\{\delta_l\}$. Using difference quotients, it is immediate to check that
 \begin{equation}\label{eq:ResDeriv}
\frac{\partial}{\partial \omega(l)}\frac{1}{H_L-z}=- \lambda\frac{1}{H_L-z}P_{l}\frac{1}{H_L-z}- g \frac{1}{H_L-z}\frac{\partial V_{\eff,\omega,L}}{\partial \omega(l)}\frac{1}{H_L-z}.
\end{equation}

Taking matrix elements we obtain from \eqref{eq:kernel} that
\begin{equation}\label{eq:melderivatives}
\frac{\partial K_L(n,\omega;t)}{\partial \omega(l)}=\sum_{m\in\Lambda_L}a(n,m)\left(-\lambda r_L(m,l;t) -g\sum_{k\in \Lambda_L}r_L(m,k;t)\frac{\partial V_{\eff,\omega,L}(k)}{\partial \omega(l)}\right)
\end{equation}
with \begin{equation}\label{eq:def r} r_L(u,v;t):=G_L(u,v;t-i\eta)G_L(v,u;t-i\eta)-G_L(u,v;t+i\eta)G_L(v,u;t+i\eta).
 \end{equation}
 The above derivatives of the kernel $K_L(n,\omega;t)$ are shown to decay exponentially in $d(n,l)$ as follows. We first rewrite $r_L(u,v;t)$ as \begin{align}\label{eq:def r2} r_L(u,v;t)=&\left(G_L(u,v;t-i\eta)-G_L(u,v;t+i\eta)\right)G_L(v,u;t-i\eta)\\
 \nonumber&+G_L(u,v;t+i\eta)\left(G_L(v,u;t-i\eta)-G_L(v,u;t+i\eta)\right).
 \end{align}
For the operators studied here the Combes-Thomas bound \cite[Theorem 10.5]{A-W-B} yields
 \begin{equation}\label{eq:CTbound}
     |G_L(u,v;z)|\leq \frac{2}{\eta}e^{-\nu d(u,v)},\,\,\,\,z\in \mathbb{C}\setminus \mathbb{R}
 \end{equation}
   for all $\nu>0$ satisfying \eqref{eq:definitionnu}. 
 Moreover, by \cite[Appendix D, Lemma 3]{A-G} we have the following inequality:
 \begin{equation}\label{eq:AGlemma}|G_L(u,v;t+i\eta)-G_L(u,v;t-i\eta)|\leq 12\eta e^{-\nu d(u,v)}\langle \delta_u,{\frac{1}{(H_L-t)^2+\eta^2/2}}\delta_u\rangle^{1/2}\langle \delta_v,{\frac{1}{(H_L-t)^2+\eta^2/2}}\delta_v\rangle^{1/2}.
  \end{equation}
  We remark that the above result, as the usual Combes-Thomas bound, may also be applied to the metric $d$ instead of the usual metric of $\mathbb{Z}^d$.
  One then obtains
  \begin{equation}\label{eq:estruv}
  \abs{r_L(u,v;t)}\leq 48e^{-2\nu d(u,v)} \langle \delta_u,{\frac{1}{(H_L-t)^2+\eta^2/2}}\delta_u\rangle^{1/2}\langle \delta_v,{\frac{1}{(H_L-t)^2+\eta^2/2}}\delta_v\rangle^{1/2}.
  \end{equation}
  By the spectral measure representation and the Cauchy-Schwarz inequality, the right-hand side of \eqref{eq:estruv} can be controlled via
  \begin{equation}\label{eq:controlintegral}\int^{\infty}_{-\infty}\langle \delta_u,{\frac{1}{(H_L-t)^2+\eta^2/2}}\delta_u\rangle^{1/2}\langle \delta_v,{\frac{1}{(H_L-t)^2+\eta^2/2}}\delta_v\rangle^{1/2}\,dt\leq \frac{\sqrt{2}\pi}{\eta}.
  \end{equation}
  Therefore,
  \begin{equation}\label{eq:integralruvest}
      \frac{1}{2\pi}\int^{\infty}_{-\infty} \abs{r_L(u,v;t)f(t)}\,dt\leq
      \frac{72\sqrt{2}\norm{F}_{\infty}}{\eta}e^{-2\nu d(u,v)}.
  \end{equation}
  Keeping in mind assumption \ref{assump5} and combining \eqref{eq:inteff}, \eqref{eq:melderivatives} and \eqref{eq:integralruvest} we reach the inequality
  \begin{align}\label{eq:subharmderV}
  \abs{\frac{\partial V_{\eff,\omega,L}(n)}{\partial\omega(l)}}&\leq \frac{C_a72\sqrt{2}\norm{F}_{\infty}}{\eta}\sum_{m\in \Lambda_L}\lambda e^{-\gamma_a d(n,m)-2\nu d(m,l)}\\
  &+\frac{C_a72\sqrt{2}\norm{F}_{\infty}}{\eta}|g|\sum_{k\in \Lambda_L}e^{-\gamma_a d(n,m)-2\nu d(m,k)}\abs{\frac{\partial V_{\eff,\omega,L}(k)}{\partial\omega(l)}} \nonumber.
  \end{align}
  We now apply Lemma \ref{Lem:subhar} with 
  fixed $l\in \Lambda_L$ and the choices $\varphi(n)=\abs{\frac{\partial V_{\eff,\omega,L}(n)}{\partial\omega(l)}}$,
 \begin{equation}
     \psi(n)=\frac{C_a72\sqrt{2}\norm{F}_{\infty}}{\eta}\lambda\sum_{m\in \Lambda_L} e^{-\gamma_a d(n,m)-2\nu d(m,l)}
 \end{equation}
 \begin{equation}
     K(n,u)=\frac{C_a72\sqrt{2}\norm{F}_{\infty}}{\eta}|g|\sum_{m\in \Lambda_L}e^{-\gamma_a d(n,m)-2\nu d(m,u)}
 \end{equation}
 in the regime where $\norm{K}_{\infty,\infty}<1$, i.e. when
  \begin{equation}\label{eq:choice1g}
    \frac{C_a72\sqrt{2}\norm{F}_{\infty}}{\eta}|g|S_{-\gamma}S_{-2\nu}<1.
  \end{equation}
In this context, introducing the weight function $W(n)=e^{\delta d(n,l)}$ with $\delta<\min\{\gamma_a,2\nu\}$ we reach 
  \begin{equation}
  b_1:=\sum_{n\in \mathbb{Z}^d}W(n)\psi(n)\leq \frac{C_a72\sqrt{2}\norm{F}_{\infty}}{\eta}\lambda S_{\delta-\gamma}S_{\delta-2\nu} 
  \end{equation}
  and 
  \begin{equation}
     b_2=\sup_{n'\in \mathbb{Z}^d}\sum_{n\in \mathbb{Z}^d}\frac{W(n)}{W(n')}K(n,n') \leq \frac{C_a72\sqrt{2}\norm{F}_{\infty}}{\eta}|g| S_{\delta-\gamma}S_{\delta-2\nu}.
  \end{equation}
   In particular, under the more restrictive assumption
  \begin{equation}\label{eq:choice2g}
      \frac{C_a72\sqrt{2}\norm{F}_{\infty}}{\eta}|g| S_{\delta-\gamma}S_{\delta-2\nu}<\frac{1}{2}
  \end{equation}
  we find that $\frac{1}{1-b_2}\leq 2$ and thus $\varphi \leq 2b_1$,
  finishing the proof.

 Given an enumeration $n_1,\ldots,\omega(n_{\abs{\Lambda'}})$ of the points in $\Lambda'$,
 it readily follows that within the smallness regime described in Lemma \ref{lem:derivativedecay}, the map
$\mathcal{T}:\mathbb{R}^{\abs{\Lambda'}}\rightarrow \mathbb{R}^{\abs{\Lambda'}}$ given by
\begin{equation}\label{eq:localdiffeo}
\mathcal{T}(\omega(n_1),\ldots,\omega(n_{\abs{\Lambda'}}))=(U^{\Lambda'}_{L}(n_1),\ldots,U^{\Lambda'}_{L}(n_{\abs{\Lambda'}})),\,\,\,
 U(n):=\omega(n)+\frac{g}{\lambda}V_{\eff,\omega,L}(n).
\end{equation} is a diffeomorphism.

We are now ready to quantify the change in $\omega$ after resampling. Fix $n_0\in \Lambda'$ and define $U^{\Lambda'}_{\alpha,L}(n)=U(n)+(\alpha-U(n_0))\delta_{n_0}$ for $n\in \Lambda'$. Then, $U^{\Lambda'}_{\alpha,L}$ is interpreted as the ``full" potential in $\Lambda'$ with value changed to $\alpha$ at $n_0$. 
Denote by $\{\omega^{\Lambda'}_{\alpha,L}(n)\}_{n\in \Lambda'}$ the random variables for which $U_{\alpha}(n)=\omega_{\alpha}(n)+V_{\eff,\omega^{\Lambda'}_{\alpha,L},L}(n)$.
In this setting we have the quasilocality result below.
  \begin{lem}\label{lem:quasilocalityomega}
  Let $C_1$ be as in \eqref{eq:C1}. Whenever
  $b_2=\frac{|g|}{\lambda}C_1<1/2$ and $\delta<\min\{\gamma_a,2\nu\}$ we have
  \begin{equation}
  \sum_{n\in \Lambda'\setminus\{n_0\}} e^{\delta d(n,n_0)}\abs{\omega^{\Lambda'}_{\alpha,L}(n)-\omega(n)}\leq \frac{2|g|C_1}{\lambda}\left(|\alpha-U(n_0)|+2\frac{|g|\norm{V_{\eff}}_{\infty}}{\lambda}\right).
  \end{equation}
  \end{lem}

\textbf{Proof.} For simplicity we denote $\omega^{\Lambda'}_{\alpha,L}$ by $\omega_{\alpha}$ in this proof. Observe that there exists $\hat \omega_{\alpha}=\{\hat \omega_{\alpha}\}_{n\in \Lambda'}$ with $\hat \omega_{\alpha}(n)\in (\omega(n),\omega_{\alpha}(n))$ such that for each $n\in \Lambda'\setminus\{n_0\}$.
\begin{align*}
\abs{\omega_{\alpha}(n)-\omega(n)}&=\frac{|g|}{\lambda}\abs{V_{\eff,\omega_{\alpha},L}(n)-V_{\eff,\omega,L}(n)}\\
&\leq \frac{|g|}{\lambda}\abs{\frac{\partial V_{\eff}(n,\hat\omega_{\alpha})}{\partial \omega(n_0)}}\left(|\alpha-U(n_0)|+\frac{|g|}{\lambda}\abs{V_{\eff,\omega_{\alpha},L}(n_0)-V_{\eff,\omega,L}(n_0)}\right)\\
&+\sum_{l\in \Lambda'\setminus\{n_0\}} \frac{|g|}{\lambda}\abs{\frac{\partial V_{\eff}(n,\hat\omega_{\alpha})}{\partial \omega(l)}}\abs{\omega_{\alpha}(l)-\omega(l)}.
\end{align*}
 Thanks to Lemma \ref{lem:derivativedecay}, whenever $\frac{\abs{g}C_1}{\lambda}<\frac{1}{2}$ we can apply Lemma \ref{Lem:subhar} with the choices
 $\varphi(n)=\abs{\omega_{\alpha}(n)-\omega(n)}\delta_{n\neq n_0}$, $W(n)=e^{\delta d(n,n_0)}$, \begin{equation}\psi(n)=\frac{|g|}{\lambda}\abs{\frac{\partial V_{\eff}(n,\hat\omega_{\alpha})}{\partial \omega(n_0)}}\left(|\alpha-U(n_0)|+\frac{|g|}{\lambda}\abs{V_{\eff,\omega_{\alpha},L}(n_0)-V_{\eff,\omega,L}(n_0)}\right)
 \end{equation}
 and
 \begin{equation}
     K(n,u)=\frac{|g|}{\lambda}\abs{\frac{\partial V_{\eff}(n,\hat\omega_{\alpha})}{\partial \omega(u)}},
 \end{equation}
 finishing the proof.

   \section{Second order decay bounds on the effective potential} \label{Sec:canceldecay}
   This section is devoted to the cancellation bounds of Step 3 of the proof outline. From now on throughout the paper we denote by $\nu$ any positive number satisfying \eqref{eq:definitionnu}.

\begin{lem}\label{lem:cancelfirstderdecay}
Let $L\in \mathbb{N}$. Whenever $\abs{g} C_a\frac{72\sqrt{2}\norm{F}_{\infty}}{\eta}S_{\frac{\delta-\delta_0}{2}}S_{-\frac{\delta_0}{2}}<\frac{1}{2}$ we have,
for each $\delta<\delta_0:=\min\{\nu,\gamma_a\}$ and $l\in \Lambda_L$,
    \begin{equation}\frac{\abs{g}}{\lambda}\sum_{n\in \Lambda_L}e^{\frac{\delta}{2} d(n,l)}\abs{\frac{\partial V_{\eff,\omega,L}(n)}{\partial \omega(l)}-\frac{\partial V_{\eff,\omega_{\alpha},L}(n)}{\partial \omega(l)}}\leq C_2\abs{\alpha- U(n_0)} 
 e^{-\frac{\delta}{2}d(n_0,l)}
    \end{equation}
    with 
    \begin{equation} \label{eq:defC2}C_2= \frac{48\norm{F}_{\infty}C_{a}}{\eta^2}S_{\frac{\delta-\delta_0}{2}}S_{-\nu}(\lambda|g|
+|g|^2C_1 )
\end{equation}
and $C_1$ as in \eqref{eq:C1}.
\end{lem}
\textbf{Proof.}

   By \eqref{eq:melderivatives} we find that if $n,l\in \Lambda_L$
\begin{align*}
\frac{\partial V_{\eff,\omega,L}(n)}{\partial \omega(l)}-\frac{\partial V_{\eff,\omega_{\alpha},L}(n)}{\partial \omega(l)}&=-\lambda\sum_{m\in\Lambda_L}a(n,m) (r_{L}(m,l)-r^{\alpha}_{L}(m,l))\\
&-g\sum_{m\in\Lambda_L}a(n,m)\sum_{k\in \Lambda_L}(r_{L}(m,k)-r^{\alpha}_{L}(m,k))\frac{\partial V_{\eff,\omega,L}(k)}{\partial \omega(l)} \\&-g\sum_{m\in\Lambda_L}a(n,m)\sum_{k\in \Lambda_L}r^{\alpha}_{L}(m,k)\left(\frac{\partial V_{\eff,\omega,L}(k)}{\partial \omega(l)}-\frac{\partial V_{\eff,\omega_{\alpha},L}(k)}{{\partial \omega(l)}}\right)
\end{align*}
where 
\begin{equation}r_L(u,v)=\frac{1}{2\pi i}\int^{\infty}_{-\infty}r_{L}(u,v;t)f(t)\,dt,
\end{equation}
$r_{L}(u,v;t)$ as in \eqref{eq:def r2} and $r^{\alpha}_{L}(u,v)$ similarly defined with $\omega$ replaced by $\omega_{\alpha,L}$.

With these definitions, letting $z=t-i\eta$ we reach
\begin{align*}
\abs{r_{L}(m,k;t)-r^{\alpha}_{L}(m,k;t)}&\leq \abs{G_{L}(m,k;z)-G^{\alpha}_{L}(m,k;z)}\abs{G_{L}(k,m;z)}\\
&+\abs{G_{L}(k,m;z)-G^{\alpha}_{L}(k,m;z)}\abs{G^{\alpha}_{L}(m,k;z)}\\
&+\abs{G_{L}(m,k;\bar{z})-G^{\alpha}_{L}(m,k;\bar{z})}\abs{G_{L}(k,m;\bar{z})}\\
&+\abs{G_{L}(k,m;\bar{z})-G^{\alpha}_{L}(k,m;\bar{z})}\abs{G^{\alpha}_{L}(m,k;\bar{z})}.\\    
\end{align*}
Note that by definition of $\omega_{\alpha}$ we have that 
\begin{equation}\label{eq:rankonecancelGreen}\abs{G_{L}(m,k;z)-G^{\alpha}_{L}(m,k;z)}=\lambda\abs{\alpha- U(n_0)}\abs{G_{L}(m,n_0;z)}{\abs{G^{\alpha}_{L}(n_0,k;z)}}
\end{equation}
for all $m,k\in \Lambda_{L}$. In particular
\begin{equation}\label{eq:cancellationrra}
    \abs{r_{L}(m,l)-r^{\alpha}_{L}(m,l)}\leq \lambda\abs{\alpha- U(n_0)} \frac{24\sqrt{2}\norm{F}_{\infty}}{\eta^2}
    e^{-\nu\left(d(m,l)+d(m,n_0)+d(n_0,l)\right)}.
    \end{equation}
    Indeed, \eqref{eq:cancellationrra} follows from \eqref{eq:rankonecancelGreen} and a similar argument to the one in \eqref{eq:controlintegral} with the help of the following Combes-Thomas type bound cf. \cite[Lemma 18]{M-S}
    \begin{equation}|G_{L}(u,v;t\pm i\eta)|\leq \sqrt{2}\langle \delta_v,\frac{1}{(H_{L}-t)^2+\eta^2/2}\delta_v\rangle^{1/2}e^{-\nu d(u,v)}
 \end{equation} 
 applied separately to $\abs{G_{L}(m,n_0;z)}$ and ${\abs{G^{\alpha}_{L}(n_0,k;z)}}$.

    Thus, 
 assumption \ref{assump1}, \eqref{eq:integralruvest} and \eqref{eq:cancellationrra} imply
\begin{align*}
&\abs{\frac{\partial V_{\eff,\omega,L}(n)}{\partial \omega(l)}-\frac{\partial V_{\eff,\omega_{\alpha},L}(n)}{\partial \omega(l)}}\leq 
\lambda^2\abs{\alpha- U(n_0)} \frac{24\sqrt{2}\norm{F}_{\infty}C_{a}}{\eta^2}e^{-\nu d(n_0,l)}\sum_{m\in\mathbb{Z}^d}e^{-\gamma d(m,n)-\nu (d(m,l)+ d(m,n_0))}\\ 
&+\abs{g}\lambda\abs{\alpha- U(n_0)} \frac{24\sqrt{2}\norm{F}_{\infty}C_{a}}{\eta^2}\sum_{m\in\mathbb{Z}^d}\sum_{k\in \mathbb{Z}^d}
    e^{-\gamma d(m,n)-\nu\left(d(m,k)+d(m,n_0)+d(n_0,k)\right)}\abs{ \frac{\partial V_{\eff,\omega,L}(k)}{\partial \omega(l)}} \\&+\abs{g}C_a\frac{72\sqrt{2}\norm{F}_{\infty}}{\eta}\sum_{m\in\mathbb{Z}^d}\sum_{k\in \mathbb{Z}^d}e^{-\gamma d(m,n)-2\nu d(m,k)}\abs{\frac{\partial V_{\eff,\omega,L}(k)}{\partial \omega(l)}-\frac{\partial V_{\eff,\omega_{\alpha},L}(k)}{{\partial \omega(l)}}}.
\end{align*}
Thus, if $\delta_0=\min\{\nu,\gamma_a\}$, $\delta<\delta_0$ and $C_1$ is as in \eqref{eq:C1}
\begin{align*}
&\abs{\frac{\partial V_{\eff,\omega,L}(n)}{\partial \omega(l)}-\frac{\partial V_{\eff,\omega_{\alpha},L}(n)}{\partial \omega(l)}}\leq 
\lambda^2\abs{\alpha- U(n_0)} \frac{24\sqrt{2}\norm{F}_{\infty}C_{a}}{\eta^2}e^{-\nu d(n_0,l)}e^{-\delta_0d(n,l)}S_{-\nu}\\ 
&+\abs{g}\lambda\abs{\alpha- U(n_0)} \frac{24\sqrt{2}\norm{F}_{\infty}C_{a}}{\eta^2}C_1S_{-\frac{\delta_0}{2}}e^{-\frac{\delta}{2}(d(n_0,l)+d(n,l))} \\&+\abs{g}C_a\frac{72\sqrt{2}\norm{F}_{\infty}}{\eta}S_{-\frac{\delta_0}{2}}\sum_{k\in \mathbb{Z}^d}e^{-\frac{\delta_0}{2}d(n,k)}\abs{\frac{\partial V_{\eff,\omega,L}(k)}{\partial \omega(l)}-\frac{\partial V_{\eff,\omega_{\alpha},L}(k)}{{\partial \omega(l)}}}.
\end{align*}

In particular, if $b_2=\abs{g} C_a\frac{72\sqrt{2}\norm{F}_{\infty}}{\eta}S_{\frac{\delta-\delta_0}{2}}S_{-\frac{\delta_0}{2}}<\frac{1}{2}$  another application of Lemma \ref{Lem:subhar} yields
\begin{align*}
&\sum_{n\in \mathbb{Z}^d}e^{\frac{\delta}{2} d(n,l)}\abs{\frac{\partial V_{\eff,\omega,L}(n)}{\partial \omega(l)}-\frac{\partial V_{\eff,\omega_{\alpha},L}(n)}{\partial \omega(l)}}\\&\leq \lambda^2\abs{\alpha- U(n_0)} \frac{48\norm{F}_{\infty}C_{a}}{\eta^2}e^{-\nu d(n_0,l)}S_{\frac{\delta-\delta_0}{2}}S_{-\nu}
\\
&+|g|\frac{48\norm{F}_{\infty}C_{a}}{\eta^2}\lambda\abs{\alpha- U(n_0)}e^{-\frac{\delta}{2}d(n_0,l)} S_{-\frac{\delta_0}{2}}S_{\frac{\delta_0-\delta}{2}}C_1\\
&\leq \lambda\abs{\alpha- U(n_0)} \frac{48\norm{F}_{\infty}C_{a}}{\eta^2}S_{\frac{\delta-\delta_0}{2}}S_{-\nu}(\lambda
+|g|C_1 )e^{-\frac{\delta}{2}d(n_0,l)}
\end{align*}
with $C_1$ as in \eqref{eq:C1}.

    \section{A pair of technical lemmas}\label{Sec:Lemmas}
    Fix $L\in\mathbb{N}$ and $\Lambda'\subset \Lambda_{L}$. Recall that in \eqref{eq:localdiffeo}
we have denoted $U(n)=\omega(n)+\frac{g}{\lambda}V_{\eff,\omega,L}(n)$ for each $n\in \Lambda'$ with $V_{\eff,\omega,L}$ given by \eqref{finitevoleffpot}. We also write $\mathcal{T}:\mathbb{R}^{|\Lambda'|} \rightarrow \mathbb{R}^{|\Lambda'|}$ the above change of variables, i.e
\begin{equation}\label{DefT} \mathcal{T}(\omega(n_1),\ldots,\omega(n_{|\Lambda'|}))=(U(n_1),\ldots,U(n_{|\Lambda'|})).
\end{equation}
In the sequel we will abbreviate this by writing 
\begin{equation*}\mathcal{T}\omega=U\,\,\text{or}\,\,\omega={\mathcal{T}}^{-1} U.
\end{equation*}

The first result on uniform control of the conditional density of $U(n_0)$ is given below.
\begin{lem}\label{lem:linfinitycontrol}
Under assumptions \ref{assump1}-\ref{assump6} whenever $\abs{g}C_a\frac{72\sqrt{2}\norm{F}_{\infty}}{\eta}S_{\delta-\gamma}S_{\delta-\nu}<\frac{1}{2}$ for some $\delta<\delta_0:=\min\{\gamma_a,\nu\}$ the conditional distribution of $U(n_0)=v$ at specified values of $\{U(n)\}_{n\in \Lambda'\setminus\{n_0\}}$
has a density $\rho^{\eff,\Lambda'}_{n_0,L}(v)$. Moreover, $\rho^{\eff,\Lambda'}_{n_0,L}(v)$ is bounded:
\begin{equation}M_{\infty}:= \sup_{\omega \in \Omega}\sup_{L\in \mathbb{N}}\sup_{\Lambda' \subset \Lambda_{L}}\sup_{n_0\in \Lambda}\sup_{v\in \mathbb{R}}\rho^{\eff,\Lambda'}_{n_0,L}(v)<\infty.
\end{equation}

\end{lem}
 
    \textbf{Proof.}
We note that in the above setting $\rho^{\eff,\Lambda'}_{n_0,L}$ is given by
    \begin{equation}\label{cdensU}\rho^{\eff,\Lambda'}_{n_0,L}(v)=\frac{ \rho\left(v-\frac{g}{\lambda}V_{\eff,{\mathcal{T}}^{-1}U,L}(n_0)\right)\prod_{n\in \Lambda'\setminus \{n_0\}}\rho\left(U(n)-\frac{g}{\lambda}V_{\eff,{\mathcal{T}}^{-1}U,L}(n)\right) J_U}{\int^{\infty}_{-\infty} \rho\left(\alpha-\frac{g}{\lambda}V_{\eff,{\mathcal{T}}^{-1}U_{\alpha},L}(n_0)\right)\prod_{n\in \Lambda'\setminus \{n_0\}}\rho\left(U^{\alpha}(n)-\frac{g}{\lambda}V_{\eff,{\mathcal{T}}^{-1}U_{\alpha},L}(n) \right) J_{U_{\alpha}}\,d\alpha} \end{equation}
Where
\begin{equation}\label{eq:jacobiansU}
    J_U=\det\Big{(}I-\frac{g}{\lambda}\frac{\partial V_{\eff,{\mathcal{T}}^{-1}U,L}(n_i)}{\partial U(n_j)}\Big{)}_{|\Lambda'|\times |\Lambda'|}\,\,\,
    J_{U_{\alpha}}=\det\Big{(}I-\frac{g}{\lambda}\frac{\partial V_{\eff,{\mathcal{T}}^{-1}U_{\alpha},L}(n_i)}{\partial U(n_j)}\Big{)}_{|\Lambda'|\times |\Lambda'|}
\end{equation}

and we recall that 
 $$U^{\alpha}(n):=U(n)+\left(\alpha-U(n_0)\right)\delta_{n=n_0}.$$
Letting $A=-\frac{g}{\lambda}\left(\frac{\partial V_{\eff,\omega,L}(n_i)}{\partial \omega(n_j)}\right)_{|\Lambda'|\times |\Lambda'|}$ and $B=-\frac{g}{\lambda}\left(\frac{\partial V_{\eff,\omega_{\alpha},L}(n_i)}{\partial \omega(n_j)}\right)_{|\Lambda'|\times |\Lambda'|}$
 one has that

\begin{equation}\label{eq:detratio}
e^{-\sum_{m,n \in \Lambda'}\abs{\left((A-B)(I+B)^{-1}\right)(m,n)}}\leq \abs{\frac{\det(I+B)}{\det(I+A)}}\leq e^{\sum_{m,n \in \Lambda'}\abs{\left((B-A)(I+A)^{-1}\right)(m,n)}}.
\end{equation}
Indeed, \eqref{eq:detratio} follows from the inequality $\det(I+M)\leq e^{\norm{M}_{1}}$ (c.f \cite[Lemma 3.3]{Simon-trace}), see  \cite[Lemma 22]{M-S}. We remark that it suffices to control ratios of the above determinants instead of the ones in \eqref{eq:jacobiansU} since the later arise from the inverse change of variables ${\mathcal{T}}^{-1}U= \omega$. 

 We are now ready to estimate the right-hand side of \eqref{eq:detratio}. Using Lemma \ref{lem:derivativedecay} we see that whenever $\frac{|g|}{\lambda}C_1<\frac{1}{4}$ we have that
 \begin{equation}
   \norm{B}_{\infty,\infty}:=\sup_{n\in \Lambda'}\sum_{l\in \Lambda'}e^{\delta d(n,l)}\abs{B(n,l)}<\frac{1}{4} 
 \end{equation}
 thus
 \begin{equation}\label{eq:CTforB}
     \abs{(I+B)^{-1}(n,l)}<4e^{-\delta d(n,l)}
 \end{equation}
by the Combes-Thomas bound. Using Lemma \ref{lem:cancelfirstderdecay} and the inequalities \eqref{eq:detratio} and \eqref{eq:CTforB} we find that  
\begin{equation}\label{eq:jacobianlowerbound}
e^{-4C_2S^2_{-\frac{\delta}{2}}|\alpha-U(n_0)|}\leq \frac{\det J_{U_{\alpha}}}{\det J_U}\leq e^{4C_2S^2_{-\frac{\delta}{2}}|\alpha-U(n_0)|}.
\end{equation}
For each $n\neq n_0$, writing $\omega_{\alpha}(n)=U^{\alpha}(n)-\frac{g}{\lambda}V^{\alpha}_{\eff,\omega,L}(n)$, one concludes from assumption \ref{assump6} that
 \begin{equation} e^{-c_{1}\abs{\omega_{\alpha}(n)-\omega(n)}}\leq \frac{\rho(\omega_{\alpha}(n))}{\rho(\omega(n))}\leq e^{c_{1}\abs{\omega_{\alpha}(n)-\omega(n)}}.
 \end{equation} By Lemma \ref{lem:quasilocalityomega} it then follows that for $\delta<\min\{\gamma_a,\nu\}$
 \begin{equation}\label{eq:productlowerbound}
     e^{-2\frac{\abs{g}}{\lambda}c_1C_1\left((|\alpha-U(n_0)|+2\frac{\abs{g}}{\lambda}\norm{V_{\eff}}_{\infty}\right)}\leq\prod_{n\neq n_0}\frac{\rho(\omega_{\alpha}(n))}{\rho(\omega(n))} \leq e^{2\frac{\abs{g}}{\lambda}c_1C_1\left((|\alpha-U(n_0)|+2\frac{\abs{g}}{\lambda}\norm{V_{\eff}}_{\infty}\right)}.
 \end{equation}
 In particular, under assumptions \ref{assump1}-\ref{assump6} for each fixed $\lambda$ we obtain for
 $\abs{g}\norm{F}_{\infty}$  sufficiently small that if
 \begin{equation}\vartheta=(2c_1\frac{\abs{g}}{\lambda}+4c_1C_1\frac{\abs{g}^2}{\lambda^2})S_{-\gamma_a}18\sqrt{2}\norm{F}_{\infty}
 \end{equation} then
 \begin{equation}\label{eq:conditionaldensityisbbdd}
   \sup_{v\in \mathbb{R}}\rho^{\eff,\Lambda'}_{n_0,L}(v)\leq e^{\vartheta}
    \sup_{v\in \mathbb{R}}\frac{\rho(v)}{\int^{\infty}_{-\infty}\rho(\alpha)e^{-\varepsilon_1\abs{v-\alpha}}\,d\alpha}<\infty,
 \end{equation}
 finishing the proof of Lemma \ref{lem:linfinitycontrol}.

Now we shall see that under assumption \ref{assump7} one may achieve a better control on the conditional densities.
 
\begin{lem}\label{lem:expdecaycontrol}
Under assumptions \ref{assump1}-\ref{assump7} there exits $\varepsilon>0$, $\vartheta=\vartheta(\norm{F},g,\lambda,\eta_0,\gamma_a,\nu,\rho)$ and $g_1=g_1(\lambda,c_1,\gamma,\nu,\eta_0)$ independent of $\Lambda'$ and $L$ such that if $\abs{g}\norm{F}_{\infty}<g_1$ then
\begin{enumerate}[label=(\roman*)]
\item \label{item:bettercontrol}
\begin{equation}
e^{-\vartheta}\left(\frac {c_{\rho}-\varepsilon}{2}\right)e^{-c_{\rho}-\varepsilon)|v|}\leq \rho^{\eff,\Lambda'}_{n_0,L}(v)\leq e^{\vartheta} \left(\frac{c_{\rho}-\varepsilon}{2} \right) e^{(-c_{\rho}+\varepsilon)\abs{v}}.
\end{equation}

\item\label{item:ratio}
\begin{equation}\label{eq:conditionalratio}
e^{-c_1(1-\vartheta)\abs{v-v'}}\leq \frac{\rho^{\eff,\Lambda'}_{n_0,L}(v)}{\rho^{\eff,\Lambda'}_{n_0,L}(v')} \leq e^{c_1(1+\vartheta)\abs{v-v'}}
\end{equation}
\end{enumerate}
Moreover, $\vartheta \to 0$ as $\abs{g}\norm{F}_{\infty}\to 0$.

\end{lem}

 \textbf{Proof.}
 To reach the upper bound we follow most of the proof of Lemma \ref{lem:linfinitycontrol}, obtaining improvements at the very end with help of assumption \ref{assump7}. Observe that, with the choice $\delta<\min\{\gamma_a,\nu\}$, equations \eqref{cdensU}-\eqref{eq:jacobianlowerbound} imply the pointwise bound
 \begin{equation}\label{eq:pointwiseupperbound}
 \rho^{\eff,\Lambda'}_{n_0,L}(v)\leq \frac{\rho(v-\frac{g}{\lambda}V_{\eff,\mathcal{T}^{-1}U,L}(n_0))}{\int^{\infty}_{-\infty} \rho(\alpha-\frac{g}{\lambda}V_{\eff,\mathcal{T}^{-1}U_{\alpha},L})e^{-2\frac{\abs{g}}{\lambda}c_1C_1\left((|\alpha-v|+2\frac{\abs{g}}{\lambda}\norm{V_{\eff,\mathcal{T}^{-1}U,L}}_{\infty}\right)}e^{-4C_2S^2_{-\frac{\delta}{2}}|\alpha-v|}\,d\alpha}
 \end{equation}
 where we recall that $C_1$ is given in \eqref{eq:C1} and is independent of $\abs{g}$. The constant $C_2$ is given in \eqref{eq:defC2} and is proportional to $\abs{g}$ when this number is sufficiently small. Note that by assumption \ref{assump6} we have for any $t\in \mathbb{R}$ and $n_0\in \Lambda'$:
 \begin{equation}
 e^{-c_1\frac{\abs{g}}{\lambda}\norm{V_{\eff,\mathcal{T}^{-1}U,L}}_{\infty}}\leq \frac{\rho(t-\frac{g}{\lambda}V_{\eff,\mathcal{T}^{-1}U,L}(n_0))}{\rho(t)}\leq e^{c_1\frac{\abs{g}}{\lambda}\norm{V_{\eff,\mathcal{T}^{-1}U,L}}_{\infty}}.
\end{equation}
Hence from \eqref{eq:pointwiseupperbound}
\begin{equation}\label{eq:2ndpointwiseupperbound}
 \rho^{\eff,\Lambda'}_{n_0,L}(v)\leq e^{2c_1(1+C_1\frac{2\abs{g}}{\lambda})\frac{\abs{g}}{\lambda}\norm{V_{\eff,\mathcal{T}^{-1}U,L}}_{\infty}}\frac{\rho(v)}{\int^{\infty}_{-\infty} \rho(\alpha)e^{-\theta|\alpha-U(n_0)|}\,d\alpha}
 \end{equation}
 with $\theta=2\frac{\abs{g}}{\lambda}c_1C_1+4C_2S^2_{-\frac{\delta}{2}}$. Now we make use of assumption \ref{assump7} to write $\frac{\rho(v)}{\rho(\alpha)}=\frac{h(v)}{h(\alpha)}e^{-c_{\rho}(|v|-|\alpha|)}$ with
 \begin{equation}\label{eq:hcomparison}
   e^{-\varepsilon_2\abs{v-\alpha}}\leq \frac{h(v)}{h(\alpha)}\leq e^{\varepsilon_2\abs{v-\alpha}}
 \end{equation} 
 and observe that \begin{equation}\label{eq:AGinfinitybound}
 \norm{V_{\eff,\omega,L}}_{\infty}\leq S_{-\gamma_a}18\sqrt{2}\norm{F}_{\infty}
 \end{equation} c.f. Theorem 3 in \cite{A-G} and assumption \ref{assump5}. This yields, with $\vartheta=(2c_1\frac{\abs{g}}{\lambda}+4c_1C_1\frac{\abs{g}^2}{\lambda^2})S_{-\gamma_a}18\sqrt{2}\norm{F}_{\infty}$,
 \begin{equation}\label{eq:3rdpointwiseupperbound}
 \rho^{\eff,\Lambda'}_{n_0,L}(v)\leq e^{\vartheta}\frac{e^{-c_{\rho}\abs{v}}}{\int^{\infty}_{-\infty}e^{-c_{\rho}\abs{\alpha}}e^{-(\varepsilon_2+\theta)\abs{v-\alpha}} \,d\alpha}.
 \end{equation}

 Pick $g_1$ sufficiently small such that if $\abs{g}\norm{F}_{\infty}<g_1$
then $\theta<\frac{\varepsilon_2}{2}$
 \begin{equation}
  \rho^{\eff,\Lambda'}_{n_0,L}(v)\leq e^{\vartheta}
    \frac{e^{-c_{\rho}\abs{v}}}{\int^{\infty}_{-\infty}e^{-c_{\rho}\abs{\alpha}}e^{-\frac{3\varepsilon_2}{2}\abs{v-\alpha}}\,d\alpha}.
 \end{equation}

 from which we readily obtain, for $\varepsilon:=\frac{3\varepsilon_2}{2}$ and 
 \begin{equation}
   \rho^{\eff,\Lambda'}_{n_0,L}(v)\leq e^{\vartheta}
  \left(\frac{c_{\rho}-\varepsilon}{2}\right)  e^{(-c_{\rho}+\varepsilon)\abs{v}}.
 \end{equation}
 The lower bound in \ref{item:bettercontrol} is analogous. One follows the above process using instead the upper bounds given in \eqref{eq:jacobianlowerbound} and \eqref{eq:productlowerbound} along with assumptions \ref{assump6}, \ref{assump7} and \eqref{cdensU} to reach
 \begin{equation}
   \rho^{\eff,\Lambda'}_{n_0,L}(v)\geq e^{-\vartheta}\left(
  \frac{c_{\rho}-\varepsilon}{2} \right) e^{(-c_{\rho}-\varepsilon)\abs{v}}.
 \end{equation}
 finishing the proof of \ref{item:bettercontrol}.
 To prove \ref{item:ratio} we use \eqref{cdensU} to write
 \begin{equation}\frac{\rho^{\eff,\Lambda'}_{n_0,L}(v)}{\rho^{\eff,\Lambda'}_{n_0,L}(v')}=\frac{ \rho\left(v-\frac{g}{\lambda}V_{\eff,{\mathcal{T}}^{-1}U,L}(n_0)\right)\prod_{n\in \Lambda'\setminus \{n_0\}}\rho\left(U(n)-\frac{g}{\lambda}V_{\eff,{\mathcal{T}}^{-1}U,L}(n)\right) J_U}{ \rho\left(v'-\frac{g}{\lambda}V_{\eff,{\mathcal{T}}^{-1}U_{v'},L}(n_0)\right)\prod_{n\in \Lambda'\setminus \{n_0\}}\rho\left(U_{v'}(n)-\frac{g}{\lambda}V_{\eff,{\mathcal{T}}^{-1}U_{v'},L}(n) \right) J_{U_{v'}}}
 \end{equation}
 Where $U_{v'}(n)=U(n)+(v'-U(n_0))\delta_{n_0}$ for $n\in \Lambda'$. 
 The bounds in \ref{item:ratio} then follow as above from \eqref{eq:productlowerbound} and \eqref{eq:jacobianlowerbound}, both applied to $\alpha=v'$, along with assumption \ref{assump6} and \eqref{eq:AGinfinitybound}.

    \section{Self-avoiding walks and localization: Proof of Theorem \ref{thm:largedisorder}}\label{sec:largedisproofs}
It is well known that the conclusion of Theorem \ref{thm:largedisorder} follows from the result below, see \cite[Appendix B]{A-S-F-H}.
\begin{thm}\label{thm:Greendecaylargedis} There exist $\lambda_{HF}$ and $g_0=g_0(C_a,d,\rho,\lambda,\gamma_a,\eta_0)$ (independent of $L$ and $\Lambda'$) such that whenever  $\lambda>\lambda_{HF}$ and $\abs{g}\norm{F}_{\infty}<g_0$ we have that for each $s\in (0,1)$
\begin{equation}\label{eq:Greendecayld}
\mathbb{E}\left(\abs{G^{\Lambda'}_{L}(m,n;z)}^s\right)\leq C_se^{-\xi_s|m-n|} 
\end{equation}
for all $z\in \mathbb{C}\setminus\mathbb{R}$ and certain constants $C_s>0$ and $\xi_s>0$ independent of $L$ and $\Lambda'$.
Moreover, $\lambda_{HF}$ solves \eqref{eq:transc equation}.
\end{thm}

\textbf{Proof.}

We closely follow the arguments of \cite{Schenkl} but provide details for the sake of completeness since a few modifications are required to account for the Hartree-Fock setting. Let $z\in \mathbb{C}\setminus \mathbb{R}$.
    We start from the depleted resolvent identity which is
    valid for $m\neq n\in \Lambda'$:
    \begin{equation}\label{eq:depletedres}
    G^{\Lambda'}_{L}(m,n;z)=-G^{\Lambda'}_{L}(m,m;z)\sum_{
    \substack{m'\in \Lambda'\\|m'-m|=1}} G^{\Lambda'\setminus\{m\}}_{L}(m',n;z).
    \end{equation}
    Note that by Lemma \ref{lem:linfinitycontrol} we have the local fractional moment bound
\begin{equation}\label{eq:aprioris}
\mathbb{E}_{U(m_j)}\left(\abs{G^{\Lambda''}_{L}(m_j,m_j;z)}^s\right)\leq \frac{(2M_{\infty})^s}{(1-s)\lambda^s}
\end{equation}   
 which is
valid for any $\Lambda''\subset \Lambda_L$ and $s\in (0,1)$, see \cite[Theorem 8.1]{A-W-B}.
    Iterating \eqref{eq:depletedres} along a sequence $m_0=m,m_1,\ldots,m_j$ of distinct points in $\Lambda'$ and applying \eqref{eq:aprioris} we find that after $N$ iterations
\begin{align*}
  \mathbb{E}\left(\abs{G^{\Lambda'}_{L}(m,n;z)}^s\right)\leq &  \sum^{N}_{j=0}\left(\frac{(2M_{\infty})^s}{(1-s)\lambda^s}\right)^j \sum_{
    \{m_k\}^{j}_{k=1}\in S^{\Lambda'}_{j}(n,m)} \mathbb{E}\left(\abs{G^{\Lambda' \setminus\{m_0,\ldots,m_j\}}_{L}(n,n;z)}^s\right) \\
    &+ \left(\frac{(2M_{\infty})^s}{(1-s)\lambda^s}\right)^N \sum_{\substack{\{m_k\}^{N}_{k=1}\in S^{\Lambda'}_{N}(m)\\ m_k\neq n\,\,k=1,\ldots,N}} \mathbb{E}\left(\abs{G^{\Lambda'\setminus \{m_0,\ldots,m_k\}}_{L}(m_k,n;z)}^s\right)
\end{align*}
where we denote by $S^{\Lambda'}_{j}(n,m)$ the set of self-avoiding walks in $\Lambda'$ of length $j$ starting at $m$ and ending at $n$ and by $S^{\Lambda'}_{N}(m)=\cup_{n\in \Lambda'}S^{\Lambda'}_{N}(n,m)$ the set of all self-avoiding walks in $\Lambda'$ of length $N$ starting at $m$. Therefore, applying \eqref{eq:aprioris} once more and denoting $\Gamma(s):=\frac{(2M_{\infty})^s}{(1-s)\lambda^s}$ we have that
\begin{equation}\label{eq:greensprovisionalbound}
  \mathbb{E}\left(\abs{G^{\Lambda'}_{L}(m,n;z)}^s\right)\leq \sum^{N}_{j=0}\Gamma(s)^{j+1} \# S^{\Lambda'}_{j}(n,m)
+ \Gamma(s)^N \# S^{\Lambda'}_{N}(m)\frac{1}{\abs{\mathrm{Im}z}^s}.
\end{equation}
We now make use of some facts about self-avoiding walks, see \cite{Schenkl} and references therein for a more detailed discussion. Recall that the self-avoiding walk correlation function is defined by
\begin{equation}
C_{\gamma}(n-m):=\sum^{\infty}_{N=0} \gamma^N\# S_{N}(n,m)
\end{equation}
whenever $\sum^{\infty}_{N=0} \abs{\gamma}^N\# S_{N}(n,m)<\infty$.
The self-avoiding walk susceptibility is defined by
\begin{equation}\label{eq:suscep}
\chi(\gamma):=\sum_{m\in \mathbb{Z}^d} C_{\gamma}(m)=\sum^{\infty}_{N=0}C_N\gamma^N
\end{equation}
where $C_N$ denotes the number of self-avoiding walks of length $N$ starting at $0$. We also recall that
the conective constant of $\mathbb{Z}^d$ is 
\begin{equation}\label{eq:conective}
    \mu_d=\lim_{N\to \infty} (C_N)^{\frac{1}{N}}.
\end{equation}
In particular, $\frac{1}{\mu_d}$ is the radius of convergence of \eqref{eq:suscep}. It is also well-known that  $0<\mu_d<2d-1$.
It is crucial for our argument that
whenever $0<\gamma<\frac{1}{\mu_d}$ the self-avoiding walk correlation function $C_{\gamma}(m)$ decays exponentially as $|m|\to \infty$. This follows from the inequality
\begin{equation}\label{eq:correlationexpbound}
 C_{\gamma}(m)\leq B_{\varepsilon} \left((\mu_d+\varepsilon)\gamma\right)^{|m|}  
\end{equation}
valid for $\varepsilon>0$ and some constant $B_{\varepsilon}$.

Therefore, whenever $\Gamma(s)<\frac{1}{\mu_d}$ we have that $\chi(\Gamma(s))\leq \sum^{\infty}_{N=0}C_N \Gamma(s)^N <\infty$. In particular, the remainder in \eqref{eq:greensprovisionalbound} satisfies
\begin{equation}
    \Gamma(s)^N \# S^{\Lambda'}_{N}(m)\leq \Gamma(s)^NC_N\to 0\,\,\,\text{as}\,\,\,N\to \infty.
\end{equation}
Thus, letting $N\to \infty$ in \eqref{eq:greensprovisionalbound} we find 
\begin{equation}
  \mathbb{E}\left(\abs{G^{\Lambda'}_{L}(m,n;z)}^s\right)\leq \sum^{\infty}_{j=0}\Gamma(s)^{j+1} \# S^{\Lambda'}_{j}(n,m).
\end{equation}
from which we conclude that
\begin{equation}
  \mathbb{E}\left(\abs{G^{\Lambda'}_{L}(m,n;z)}^s\right)\leq \Gamma(s) C_{\Gamma(s)}(m-n).\end{equation}
  Finally, to end the proof we determine for which values of $s\in (0,1)$ one has that $\Gamma(s)<\frac{1}{\mu_d}$. Observe that whenever $\frac{\lambda}{2M_{\infty}}>e$ the only critical point of $\Gamma(s)$ is $s_0(\lambda)=1-\frac{1}{\ln\left(\frac{\lambda}{2M_{\infty}}\right)}$ which yields
  \begin{equation}
  \Gamma(s_0)=e\ln \left(\frac{\lambda}{2M_{\infty}}\right)\frac{2M_{\infty}}{\lambda}.
  \end{equation}
  Thus $\Gamma(s_0)<\frac{1}{\mu_d}$ if and only if
  \begin{equation}\lambda>{2M_{\infty}}\mu_d e  \ln \left(\frac{\lambda}{2M_{\infty}}\right)
  \end{equation}
  so the critical threshold is
  $\lambda_{\mathrm{HF}}=2M_{\infty}\ln \left(\frac{\lambda_{\mathrm{HF}}}{(2M_{\infty})}\right)\mu_d e$. For values of $\lambda$ greater than $\lambda_{HF}$ we conclude that there exists $\varepsilon>0$ for which
  \begin{equation}
  \mathbb{E}\left(\abs{G^{\Lambda'}_{L}(m,n;z)}^{1-\frac{1}{\ln\left(\frac{\lambda}{2M_{\infty}}\right)}}\right)\leq 
  e\ln \left(\frac{\lambda}{2M_{\infty}}\right)\frac{2M_{\infty}}{\lambda} B_{\varepsilon}\left((\mu_d+\varepsilon)e\ln \left(\frac{\lambda}{2M_{\infty}}\right)\frac{2M_{\infty}}{\lambda}\right)^{|m-n|}.
  \end{equation}
  and $(\mu_d+\varepsilon)e\ln \left(\frac{\lambda}{2M_{\infty}}\right)\frac{2M_{\infty}}{\lambda}<1$.
  Applying H\"older's inequality we conclude that \eqref{eq:Greendecayld} holds for any $s\in (0,1)$ and some $C_s>0$ and $\xi_s>0$. This is immediate if $0<s<1-\frac{1}{\ln\left(\frac{\lambda}{2M_{\infty}}\right)}$ and follows from (off-diagonal) a-priori bounds for the Green's function if $1-\frac{1}{\ln\left(\frac{\lambda}{2M_{\infty}}\right)}<s<1$, see \cite[Lemma B2]{A-S-F-H} and \cite[Theorem 8.3]{A-W-B}.

  \section{ Proof of theorem \ref{thm:weakdisorder} and Corollary \ref{cor:stability}}
  \label{Sec:weakdisproof} 
  Similarly to how Theorem \ref{thm:Greendecaylargedis} implies Theorem \ref{thm:largedisorder}, Theorem \ref{thm:weakdisorder} follows from the result below.
  \begin{thm}\label{thm:Greendecayweakdis}
In the setting of Lemma \ref{lem:expdecaycontrol}, for each $I\subset \mathbb{R}$ the exists $g_1(C_a,d,\rho,\lambda,\gamma_a,\eta_0)$, $\nu''>0$, $C>0$ and $\lambda_0$ (independent of $\Lambda'$ and $L$) such that whenever $|g|\abs{F}_{\infty}<g_1$ and $\lambda<\lambda_0$ we have that
\begin{equation}\label{eq:Greendecaywd}
\mathbb{E}\left(\abs{G_{\Lambda}(m,n,E)}^s\right)\leq Ce^{-\nu''|m-n|} 
\end{equation}
for some $s\in(0,1)$. Moreover,
we have that \begin{equation}
    \lambda_0=\sup_{s\in (0,1)}\sup_{\mu>0}\inf_{E\in I}\widehat{\lambda}_{s,\mu}(E)
\end{equation}
where
\begin{equation}
 \widehat{\lambda}_{s,\mu}(E)=\left( D_{s,1}\sup_{\delta\neq 0}\sup_{u\in \mathbb{Z}^d}\sum_{v\in \mathbb{Z}^d}|G_0(u,v;E+i\delta)|^se^{\mu|u-v|}\right)^\frac{-1}{s}.
\end{equation}
\end{thm}
  Theorem \ref{thm:Greendecayweakdis} in turn follows from Lemma \ref{lem:expdecaycontrol} along with known results and thus we only provide an outline for how it is proven. Before doing so, we recall some notions of regularity for probability distributions, c.f. \cite{A-Molc,A-W-B} which will be relevant in the sequel.
  \begin{dfn}
\begin{enumerate}[label=(\roman*)]
\item{
A probability measure $\rho(dv)$ on the real line is $\tau$-regular, with $\tau\in(0,1]$, if for some $v_0 \in \mathbb{R}$ and $C>0$
\begin{equation}\label{eq:taureg}
    \rho\left([v-\delta,v+\delta]\right)\leq C|\delta|^{\tau}\rho\left([v-v_0,v+v_0]\right)
\end{equation}
holds for all $\delta\in (0,1)$ and $v\in \mathbb{R}$.}
\item{ A joint probability measure $\rho(dV)$ of a collection of random variables $\{V_n\}$ is conditionally $\tau$-regular if the conditional distributions of $V_{n}$ at specified values of $\{V_m\}_{m\neq n}$ satisfy \eqref{eq:taureg} with uniform values of the constants appearing there.}
\item{If, additionally, for some $\varepsilon>0$ the conditional expectations of $|V_n|^{\varepsilon}$ are uniformly bounded:
\begin{equation}\label{eq:varepsilonreg}
\mathbb{E}\left(|V_n|^{\varepsilon}|\,\,V_{\{n\}^c}\right)\leq B,\,\,\,\,\,\,\text{for some}\,\,\, B>0,
    \end{equation}
    then the joint probability measure $\rho(dV)$ is said to be conditionally $(\tau,\varepsilon)$-regular.}

    \item $\rho$ has regular $q$-decay for $q>0$ if
    \begin{equation}
        \rho\left([u-1,u+1]\right)\leq \frac{C}{1+\abs{u}^q},\,\,\,\,\,\,\text{for some}\,\,\, C>0.
    \end{equation}
    \end{enumerate}
    \end{dfn}
    \textbf{Proof of Theorem \ref{thm:Greendecayweakdis}:} Lemma \ref{lem:expdecaycontrol} \ref{item:bettercontrol} readily implies that $\rho^{\eff,\Lambda'}_{n_0,L}(v)\,dv$ has regular $q$ decay for all $q>0$ and that for all $p>0$
   $$\int^{\infty}_{-\infty}|v|^p\rho^{\eff,\Lambda'}_{n_0,L}(v)\,dv<\infty,$$
   i.e. $\rho^{\eff,\Lambda'}_{n_0,L}(v)\,dv$ is conditionally $(1,p)$-regular for all $p>0$.
   Moreover, by Lemma \ref{lem:expdecaycontrol} \ref{item:ratio}, we have that for any $\delta\in (0,1]$ and $u\in \mathbb{R}$
   \begin{align*}
   \int^{u+\delta}_{u-\delta}\rho^{\eff,\Lambda'}_{n_0,L}(v)\,dv&\leq (2\delta) e^{c_1(1+\vartheta)}
   \rho^{\eff,\Lambda'}_{n_0,L}(u)\\
   &\leq \delta e^{2c_1(1+\vartheta)}\int^{u+1}_{u-1}\rho^{\eff,\Lambda'}_{n_0,L}(v)\,dv,
   \end{align*}
   in particular we see that $\rho^{\eff,\Lambda'}_{n_0,L}(v)\,dv$ is (uniformly) $1$-regular.
   
   We then conclude from \cite[Theorem 8.7]{A-W-B} that $\rho^{\eff,\Lambda'}_{n_0,L}$ is $1$-moment regular, namely $D_{s,1}<\infty$ with $D_{s,1}$ as in \eqref{eq:decoupconst}
for all $s\in (0,1)$. In particular, Theorem \ref{thm:Greendecayweakdis} falls into the framework of \cite[Theorem 10.4]{A-W-B} .

\textbf{Proof of Corollary \ref{cor:stability}:} Note that when $\abs{g}\norm{F}_{\infty}\to 0$ then $\theta \to 0$ in equation \eqref{eq:2ndpointwiseupperbound} ( which only requires assumptions \ref{assump1}-\ref{assump6}). Thus, by dominated convergence, we may choose $M_{\infty}$ such that $M_{\infty}\to {\norm{\rho}}_{\infty} $ as $\abs{g}\norm{F}_{\infty}\to 0$. Corollary \ref{cor:stability} now follows from \eqref{eq:transced} and \eqref{eq:transced2} since these equations imply
\begin{align*}(\lambda_{\mathrm{HF}}-\lambda_{\mathrm{And}})-2M_{\infty}(\ln(\lambda_{\mathrm{HF}})-\ln(\lambda_{\mathrm{And}}))=& 2\mu_d\ln(\lambda_{\mathrm{And}})(M_{\infty}-\norm{\rho}_{\infty})\\
&2\mu_d(\norm{\rho}_{\infty}\ln(2\norm{\rho}_{\infty})-M_{\infty}\ln(2M_{\infty}))
\end{align*}
and by construction $\frac{\lambda_{\mathrm{HF}}}{2M_{\infty}}>e>1.$

\appendix

 \section{Appendix}
 We now provide some results on existence and uniqueness of the effective potentials as well as their regularity with respect to the random variables. Since the statements are mostly immediate generalizations from the ones given in \cite{M-S} we skip most proofs. We formulate the first of these results for $\ell^{\infty}\left(\mathbb{Z}^d\right)$ but remark that its finite volume analogue holds similarly.
 \subsection{Contraction mapping arguments}
  Let $\Phi:\ell^{\infty}\left(\mathbb{Z}^d\right)\rightarrow \ell^{\infty}\left(\mathbb{Z}^d\right)$
  be given by
  \begin{equation}
      \Phi(V)=\sum_{m\in \mathbb{Z}^d}a(n,m)\langle{\delta_m},F(A+\lambda V_{\omega}+gV)\delta_m\rangle
  \end{equation}
  We wish to show that there is a unique solution $V_{\eff}$ to the equation $\Phi(V)=V$. For that purpose, we introduce a technical Lemma which may be found in \cite[Proposition 12]{M-S}
  
  \begin{lem}\label{Lem:Contraction} 
\begin{enumerate}[label=(\alph*)]
\item{ Let $T=A+\lambda V_{\omega}$ be as in assumptions \ref{assump1}-\ref{assump5}. Given potentials $V,W \in \ell^{\infty}\left(\mathbb{Z}^d\right)$, we have, for any $\nu$ satisfying \eqref{eq:definitionnu} and $\delta\in (0,\nu)$, that
 \begin{equation}\label{contraction1}  \Big|\langle \delta_m,\left(F(T+V)-F(T+W)\right)\delta_n\rangle\Big|\leq \frac{72\sqrt{2}}{\eta}S_{\delta-\nu}\|F\|_{\infty}\|V-W\|_{\infty}e^{-\nu'd(m,n)}.
\end{equation}}

\item{ For any $m,n,j \in\mathbb{Z}^d$, the matrix elements $\langle \delta_m,F(T+gV)\delta_n\rangle$ are differentiable with respect to $V(j)$ and \begin{equation}\label{contraction3}\Big|\frac{\partial\langle \delta_m,F(T+gV)\delta_n \rangle}{\partial V(j)}\Big|\leq |g|\frac{72\sqrt{2}e^{-\nu\left(|d(m,j)+d(n,j)\right)}}{\eta}\|F\|_{\infty}\|V\|_{\infty}.
\end{equation}}

\end{enumerate}
\end{lem}

From Lemma \ref{Lem:Contraction} and assumption \ref{assump5} we obtain
\begin{align*}
    \norm{\Phi(V)-\Phi(W)}_{\infty}&\leq |g|\frac{72\sqrt{2}}{\eta}S_{\delta-\nu}S_{-\gamma_a}C_a\|F\|_{\infty}\|V-W\|_{\infty}
\end{align*}
thus we conclude the following.
\begin{prop}
    Whenever $|g|\frac{72\sqrt{2}}{\eta}S_{\delta-\nu}S_{-\gamma_a}C_a\|F\|_{\infty}<1$ for some $\delta\in (0,\nu)$ the map $\Phi:\ell^{\infty}\left(\mathbb{Z}^d\right)\rightarrow \ell^{\infty}\left(\mathbb{Z}^d\right)$ is a contraction. In particular, there is a unique $V_{\eff}\in \ell^{\infty}\left(\mathbb{Z}^d\right)$ such that $\Phi(V_{\eff})=V_{\eff}$. Moreover, the analogue effective potential in finite volume $\Lambda_L$, $V_{\eff,\omega,L}$, is a smooth function of $(\omega(n_1),...,\omega(n_{\abs{\Lambda_L}}))$.
\end{prop}
We also note that if $a(n,m)\in \mathbb{R}$ for each $n,m \in \mathbb{Z}^d$ then $V_{\eff}(n)\in \mathbb{R}$ for each $n \in \mathbb{Z}^d$.
\subsection{Norm resolvent convergence}
Finally, we briefly comment on the convergence of resolvents which allows to extend the results of Theorems \ref{thm:largedisorder} and \ref{thm:weakdisorder} to infinite volume operators. It will be useful to introduce the augumented boundary
 \begin{equation}
        \partial \Lambda_L=\{u\in \mathbb{Z}^d:\,\,\mathrm{dist}(u,\Lambda_L)=1\,\,\text{or}\,\,\mathrm{dist}(u,\Lambda^{c}_L)=1\}
    \end{equation}
    with $\mathrm{dist}(u,X)$ calculated in the metric of $\mathbb{Z}^d$.
\begin{lem}\label{lem:normresconv}
\begin{enumerate}[label=(\alph*)]
    \item\label{eq:a)potlocal} Given $n\in \Lambda_L$ whenever $ \frac{3\sqrt{2}\abs{g}\norm{F}_{\infty}S_{-\nu}}{\eta}<\frac{1}{2}$ we have that
    \begin{equation}
        \abs{V_{\eff,\omega}(n)-V_{\eff,\omega,L}(n)}\leq C e^{-\delta d(n,\partial \Lambda_L)}
    \end{equation}
    for any $\delta<\min\{\nu,\gamma_a\}$ and
     $C=\frac{432C_a\norm{F}_{\infty}\abs{g}S_{-\nu}}{\eta}$, with
     $d(n,\partial \Lambda_L)$ calculated in the metric $d(\cdot,\cdot)$ of assumption \ref{assump5}.

    \item \label{b) green local} 
    For any $\kappa>0$, with $\abs{g}$ and $\delta$ as above
    \begin{equation}
        \abs{G^{\Lambda_L}(m,n;t+i\kappa)-G^{\Lambda_L}_{L}(m,n;t+i\kappa)}\leq \frac{4C}{\kappa^2} e^{-\nu d(m,n)-\delta d(n,\partial \Lambda_L))}S_{-\nu}
    \end{equation}
    In particular, for each fixed $z\in \mathbb{C}^{+}$ and $\psi\in \ell^2\left(\mathbb{Z}^d\right)$ we have that
    \begin{equation}
        \norm{(H^{\Lambda_L}-z)^{-1}\psi-{(H^{\Lambda_L}_{L}-z)^{-1}\psi}}\to 0 \,\,\,\text{as}\,\,\,L\to \infty.
    \end{equation}
    
\end{enumerate}

\end{lem}

\textbf{Proof.}
Using \eqref{eq:inteff} and the analogous representation for $V_{\eff,\omega}(n)$ we find
\begin{equation*}
 \abs{V_{\eff,\omega}(n)-V_{\eff,\omega,L}(n)}\leq \frac{3\norm{F}_{\infty}}{2\pi}\int^{\infty}_{-\infty} \abs{K(n,\omega;t)-K_L(n,\omega;t)}\,dt  
\end{equation*}
where for $z=t-i\eta$
\begin{align*}\abs{K(n,\omega;t)-K_L(n,\omega;t)}&\leq \sum_{m\in \mathbb{Z}^d}\abs{a(n,m)}\abs{G(m,m;z)-G_L(m,m;z)}\\
&\sum_{m\in \mathbb{Z}^d}\abs{a(n,m)}\abs{G(m,m;\bar z)-G_L(m,m;\bar z)}
\end{align*}
Observe that letting $\Lambda_L^\mathrm{o}:=\Lambda_L\setminus\partial \Lambda_L$ and $(\Lambda_L^\mathrm{o})^c=\mathbb{Z}^d\setminus \Lambda_L^\mathrm{o}$, for any $m\in \mathbb{Z}^d$ we have that
\begin{align*}\abs{G(m,m;z)-G_L(m,m;z)}&\leq \abs{g}\sum_{k\in \Lambda_L^\mathrm{o}} \abs{G(m,k;z)}\abs{V_{\eff,\omega}(k)-V_{\eff,\omega,L}(k)}\abs{G_L(k,m;z)}\\
&36 \sqrt{2}\norm{F}_{\infty}S_{-\gamma_a}\abs{g}\sum_{k'\in (\Lambda_L^\mathrm{o})^c}\abs{G(m,k';z)}\abs{G_L(k',m;z)}\\
\end{align*}
where we have used that $\max\{\norm{V_{\eff,\omega,L}}_{\infty},\norm{V_{\eff,\omega}}_{\infty}\leq S_{-\gamma_a}18\sqrt{2}\norm{F}_{\infty}$ c.f. Theorem 3 in \cite{A-G} and assumption \ref{assump5}.

The result in \ref{eq:a)potlocal} now follows from 
\begin{equation}
 \int^{\infty}_{-\infty}
 \abs{G(u,v;z)}\abs{G_L(v,u;z)}\,dt\leq \frac{2\sqrt{2} \pi}{\eta}e^{-2\nu d(u,v)} 
\end{equation}
combined with assumption \ref{assump5} and another application of Lemma \ref{Lem:subhar} with
\begin{equation}
\varphi(n)=\abs{V_{\eff,\omega}(n)-V_{\eff,\omega,L}(n)},\,\,\, W(n)=e^{\delta d(n,{\partial\Lambda_L})} 
\end{equation} and
\begin{equation}
K(n,u)=\left(\sum_{m\in \mathbb{Z}^d} e^{-\gamma_a d(n,m)-2\nu d(m,u)}\right) \mathds{1}_{{\Lambda_{L}}^\mathrm{o}}(u)
\end{equation}
for which we have $b_1=\frac{216\norm{F}_{\infty}\abs{g}S_{-\nu}}{\eta}$ and $b_2=\frac{3\sqrt{2}\abs{g}\norm{F}_{\infty}S_{-\nu}}{\eta}$.

\ref{b) green local} now follows from \ref{eq:a)potlocal} combined with the resolvent identity and another application Combes-Thomas bound.

   \subsection*{Acknowledgements}
Dedicated to Abel Klein in ocasion of his 78th birthday. This work was partially supported by NSF DMS-2000345 and DMS-2052572. R. Matos is thankful to the anonymous reviewer for several remarks which greatly improved the exposition in this note.

\bibliographystyle{abbrv}
\bibliography{main_locregime paper}
 
\end{document}